\documentclass[aps,pra,twocolumn,showpacs,preprintnumbers,amsmath,amssymb,
floatfix,superscriptaddress]{revtex4-2}

\usepackage{graphicx}
\usepackage{dcolumn}
\usepackage{bm}
\usepackage[usenames,dvipsnames]{color} 
\usepackage[table]{xcolor} 
\usepackage{soul}
\usepackage{physics}

\usepackage[colorlinks=true, linkcolor=blue, citecolor=blue, urlcolor=blue]{hyperref}
\usepackage{comment}

\newcommand{\ybe}{\ensuremath{^{171}\text{Yb}^+\,}}
\newcommand{\ybz}{\ensuremath{^{172}\text{Yb}^+\,}}
\newcommand{\ybd}{\ensuremath{^{173}\text{Yb}^+\,}}
\newcommand{\yb}{\ensuremath{\text{Yb}^+\,}}

\newcommand{\affA}{Physikalisch-Technische Bundesanstalt, Bundesallee 100, 38116
Braunschweig, Germany}
\newcommand{\affB}{Institut f\"ur Quantenoptik, Leibniz
Universit\"at Hannover, Welfengarten 1, 30167 Hannover, Germany}
\newcommand{\affC}{National Institute of Metrology (Thailand),
 3/4-5 Moo 3, Klong 5, Klong Luang, 12120 Pathumthani, Thailand}

\begin{document}

\preprint{APS/123-QED}

\title{Nuclear spin quenching of the $^2S_{1/2}\rightarrow {^2}F_{7/2} $ electric octupole transition in $^{173}$Yb$^+$ }

\author{Jialiang Yu}\email{jialiang.yu@ptb.de}
\affiliation{\affA}
\author{Anand Prakash}\affiliation{\affB}
\author{Clara Zyskind}\affiliation{\affA}
\author{Ikbal Ahamed Biswas}\affiliation{\affA}
\author{Rattakorn Kaewuam}\affiliation{\affC}
\author{Piyaphat Phoonthong}\affiliation{\affC}
\author{Tanja E. Mehlst\"aubler}\email{tanja.mehlstaeubler@ptb.de}
\affiliation{\affA}\affiliation{\affB}



\date{\today}

\begin{abstract}
We report the coherent excitation of the highly forbidden $^2S_{1/2} \rightarrow {^2}F_{7/2}$ clock transition in the odd isotope $^{173}\mathrm{Yb}^+$ with nuclear spin $I = 5/2$, and reveal the hyperfine-state-dependent, nuclear-spin–induced quenching of this transition. The inferred lifetime of the $F_e = 4$ hyperfine state is one order of magnitude shorter than the unperturbed ${^2}F_{7/2}$ clock state of \ybe.
This reduced lifetime lowers the required optical power for coherent excitation of the clock transition, thereby reducing the AC Stark shift caused by the clock laser. Using a 3-ion Coulomb crystal, we experimentally demonstrate an approximately 20-fold suppression of the AC Stark shift, a critical improvement for the scalability of future multi-ion $\mathrm{Yb}^+$ clocks. Furthermore, we report the $|^2S_{1/2},F_g=3\rangle~\rightarrow~|^2F_{7/2},F_e=6\rangle$ unquenched reference transition frequency as $642.11917656354(43)$ THz, along with the measured hyperfine splitting and calculated quadratic Zeeman sensitivities of the ${^2}F_{7/2}$ clock state. Our results pave the way toward multi-ion optical clocks and quantum computers based on \ybd.
\end{abstract}

\maketitle




Trapped ion systems have widely been used to advance fundamental physics and develop high-precision optical clocks \cite{hausser115In+172YbCoulombCrystal2024, huntemannHighAccuracyOpticalClock2012,tofful171YbOptical2024, huntemannSingleIonAtomicClock2016a, baynhamAbsoluteFrequencyMeasurement2018,godunFrequencyRatioTwo2014a, rosenbandFrequencyRatioHg2008,kingOpticalAtomicClock2022,zhiqiang176LuClock2023,dubeSrSingleionClock2016, delehayeSingleionTransportableOptical2018, barwoodAgreementTwo882014, brewer27QuantumLogicClock2019, dorscherOpticalFrequencyRatio2021, sannerOpticalClockComparison2019,dreissenImprovedBoundsLorentz2022,hurEvidenceTwoSourceKing2022,doorProbingNewBosons2025a,onoObservationNonlinearityGeneralized2022,langeLifetime722021,filzingerImprovedLimitsCoupling2023}. In particular, the forbidden electric octupole (E3) transition in the \yb ion, with a measured lifetime of 1.6(1) years \cite{langeLifetime722021,szSurzhykovPrivComm}, offers both exquisite clock accuracy \cite{hausser115In+172YbCoulombCrystal2024, huntemannHighAccuracyOpticalClock2012, tofful171YbOptical2024, huntemannSingleIonAtomicClock2016a, baynhamAbsoluteFrequencyMeasurement2018, godunFrequencyRatioTwo2014a} and high sensitivity to new physics beyond the Standard Model, such as temporal variation of the fine-structure constant \cite{dzubaSpaceTimeVariationPhysical1999,dzubaRelativisticCorrectionsTransition2008,filzingerImprovedLimitsCoupling2023}, violations of local Lorentz invariance \cite{dzubaStronglyEnhancedEffects2016,shanivNewMethodsTesting2018a, sannerOpticalClockComparison2019,dreissenImprovedBoundsLorentz2022}, and searches for fifth forces~\cite{berengutProbingNewLongRange2018, hurEvidenceTwoSourceKing2022,doorProbingNewBosons2025a,onoObservationNonlinearityGeneralized2022}. Precision spectroscopy on \yb isotopes also opens a new window to nuclear physics \cite{dzubaAtomicElectricDipole2007,gingesTestingAtomicWave2018} and the study of deformations of neutron-rich nuclei~\cite{hurEvidenceTwoSourceKing2022,doorProbingNewBosons2025a,xiaoHyperfineStructureYb2020}. 

The sensitivity of single-ion precision spectroscopy is fundamentally limited by the quantum projection noise (QPN), which scales inversely with $\sqrt{N\cdot \tau}$, where $N$ and $\tau$ are the ion number and the measurement time, respectively. 
Thus, increasing the number of ions in clocks and precision spectroscopy can extend the reach of searches for new physics.
In the case of the \yb~E3 transition, however, scalability and multi-ion operation face challenges due to the strong AC Stark shift caused by the clock laser (hereafter referred to as the AC Stark shift).
While advanced Ramsey techniques can compensate for the AC Stark shift during clock interrogation, achieving sufficient excitation rates across multiple ions remains challenging and requires laser beam profile inhomogeneities of less than 2$\%$ \cite{yuPrecisionSpectroscopyYb2024a}.

In 2016, Dzuba and Flambaum predicted a strong hyperfine-induced electric dipole (HFE1) contribution in \ybd \cite{dzubaHyperfineInducedElectricDipole2016} due to the large electric quadrupole moment of its deformed nucleus \cite{doorProbingNewBosons2025a}.
Among stable Yb isotopes, this unique property of \ybd could effectively enhance the extremely weak $^2S_{1/2} \rightarrow {^2}F_{7/2}$ transition, which otherwise is a pure E3 transition, thereby reducing the required laser power, and thus the AC Stark shift.
The $^{173}\mathrm{Yb}$ isotope, which has also recently been proposed for studying the nuclear magnetic quadrupole moment and parity violation \cite{sunagaMeasuringNuclearMagnetic2024}, is still largely unexplored.
In addition, precise measurement of the \ybd hyperfine (HF) structure is expected to resolve higher-order nuclear moments \cite{suekaneHexadecapoleMomentsAtomic1957, sternheimerAntishieldingNuclearElectric1961}, offering insights into the nuclear structure of heavy atoms. Such investigations can help to refine theoretical models and address the discrepancy between two reported values of the magnetic octupole moment \cite{singhObservationNuclearMagnetic2013,degrooteMagneticOctupoleMoment2021}.
Overall, the reduced AC Stark shift and the narrow linewidth make \ybd an intriguing candidate for multi-ion clock spectroscopy and nuclear structure studies \cite{xiaoHyperfineStructureYb2020}.

Beyond its role in clock spectroscopy, \ybd is
promising for scalable quantum computing architectures.
In addition, \yb has been proposed for a quantum computing protocol interrogating multiple qubit types within a single atomic species \cite{allcockEmphOmgBlueprintTrapped2021,anHighFidelityState2022}.
The long-lived $^2F_{7/2}$ state, with its rich HF structure, also supports polyqubit construction \cite{campbell2022polyqubitquantumprocessing}, the implementation of advanced quantum error correction techniques and offering a new shelving transition for high detection fidelity \cite{kangQuantumErrorCorrection2023,vizvaryEliminatingQubitTypeCrossTalk2024}.
  
In this letter, we determine the absolute frequency of the $4f^{14}6s~|^2S_{1/2} ,F_g=3\rangle \rightarrow~ 4f^{13}6s^2~|^2F_{7/2},F_e=6\rangle$ clock transition, the HF splittings and the quadratic Zeeman sensitivities of the $^2F_{7/2}$ clock states, and quantify the enhancement of decay rates in the $|^2F_{7/2}, F_e = 2,4 \rangle$ hyperfine states due to nuclear spin quenching.

\begin{figure}[]
	\centering\includegraphics[width=0.48\textwidth]{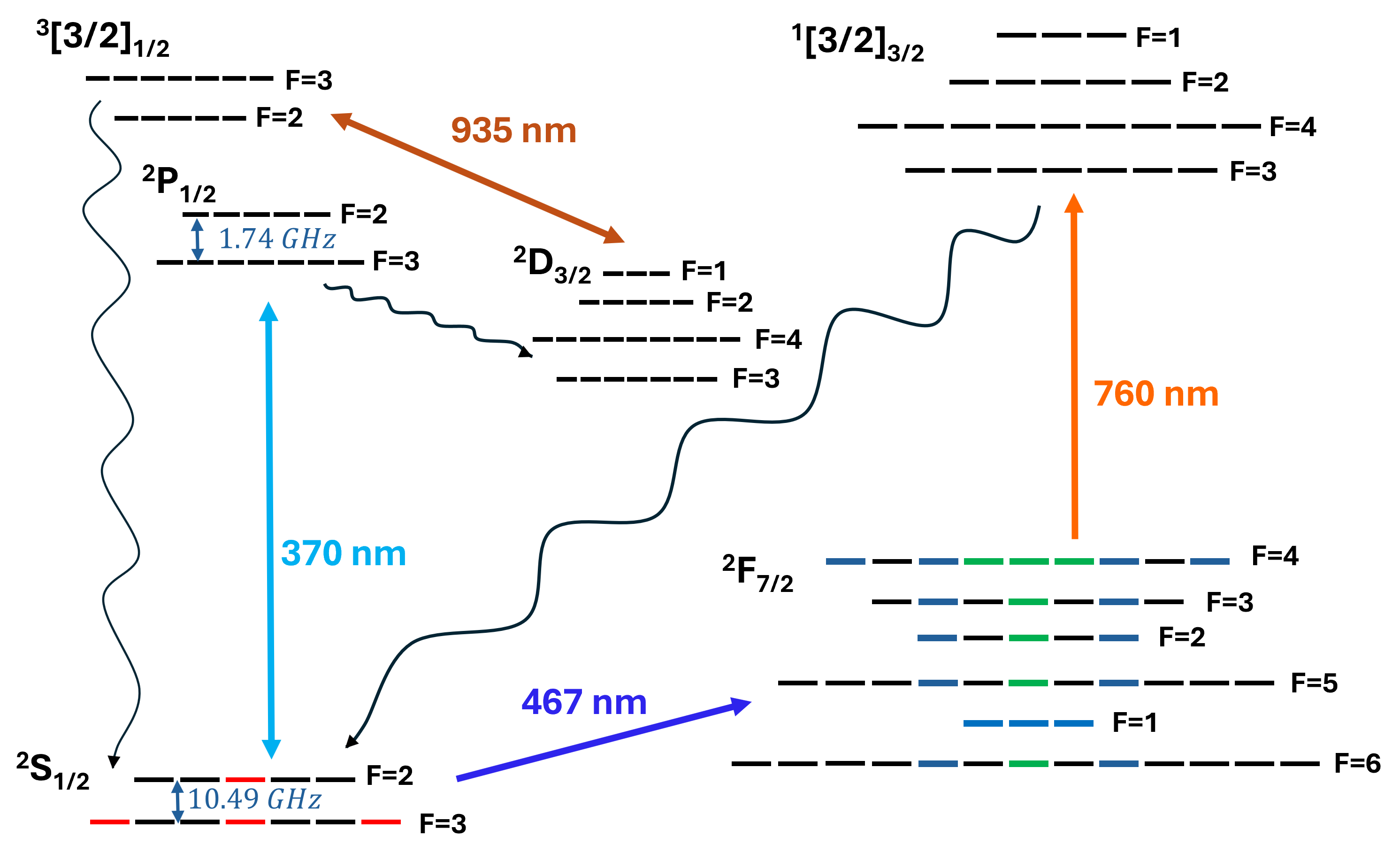}
	\caption{\label{fig:level_scheme}
		Level scheme of the $^{173}\mathrm{Yb}^{+}$ ion. The 370 nm dipole transition is used for Doppler cooling and state detection. The 935 nm laser prevents population trapping in the $^2$D$_{3/2}$ state. A 467 nm laser excites the highly forbidden electric octupole transition. Laser light at 760 nm repumps the electron from the $^2$F$_{7/2}$ state. For spectroscopic measurements, the electron population can be prepared in the $|F_g=3, m_g=0,\pm3\rangle$ or $|F_g=2, m_g=0\rangle$ (red) of the $^2S_{1/2}$ state. The interrogated $^2$F$_{7/2}$ HF states are marked in green (coherent excitation) or blue (rapid adiabatic passage).}
\end{figure}

In our setup, we initially confine a single $\mathrm{Yb}^+$ ion in a radiofrequency (RF) Paul trap \cite{kellerProbingTimeDilation2019a} with motional trapping frequencies of $\omega_x \approx 2\pi \times 650$ kHz, $\omega_y \approx 2\pi \times 612$ kHz, and $\omega_z \approx 2\pi \times 212$ kHz.
Three pairs of coils on the vacuum chamber are used to compensate for the background magnetic field and to generate a defined quantization field of $B=0.2$ mT.
Doppler cooling to temperatures below 1~mK and state detection are accomplished by 370 nm laser beams assisted by repumper lasers at 935 nm and 760 nm (Fig.~\ref{fig:level_scheme}).
To address the HF states, we use a 10.49~GHz resonant electro-optic modulator (EOM) at 370 nm and a broadband EOM at 935 nm.
As we resonantly excite the $|^2P_{1/2}, F=3\rangle$ state, and the transition from $|^2S_{1/2}, F_g=3, m=0\rangle$ to $|^2P_{1/2}, F=3, m=0\rangle$ is forbidden according to angular momentum selection rules, a $\pi$-polarized 370 nm beam is used to efficiently pump the atomic population to the $|^2S_{1/2}, F_g=3, m=0\rangle$ state.
Optionally, the atomic population can be transferred to $|^2S_{1/2}, F_g=2, m=0\rangle$ by a successive RF pulse at 10.49~GHz, resonant with the $^2S_{1/2}$ state HF splitting.
State preparation into the $|^2S_{1/2}, F_g=3, m=\pm 3\rangle$ stretched states can be achieved by utilizing $\sigma$-polarized 370 nm light.
The highly forbidden clock transitions are probed with an ultra-stable laser at 467 nm, which obtains its stability from a cryogenic silicon cavity \cite{matei15LasersSub102017}.
The clock laser, with 6 mW power, is focused to a 30 $\mathrm{\mu m}$ waist. It propagates orthogonally to the plane defined by the trap and quantization axes, with linear polarization parallel to the quantization axis.
Laser light at 760 nm repumps the electron back to the ground state after excitation.
Further details of the experimental setup can be found in the Appendix. 

To search for the $^2S_{1/2}\rightarrow {^2F_{7/2}}$ clock transitions in \ybd, we predict its unperturbed frequency (without hyperfine interaction and systematic shifts) using a King plot \cite{kingIsotopeShiftsAtomic1984, kingCommentsArticlePeculiarities1963a, doorProbingNewBosons2025a}, which combines the transition frequencies of the five even isotopes \cite{doorProbingNewBosons2025a} with the measured $^2S_{1/2} \rightarrow{^2D_{5/2}}$ quadrupole transitions for $^{173}\mathrm{Yb}^+$, yielding $\nu_{173} = 642.118148(243)$ THz.
To determine the HF structure, we extrapolate the $A$-coefficient of $^{173}\mathrm{Yb}^+$ to $A_{173} = -249.3(3)$ MHz from the known $A$-coefficient of $^{171}\mathrm{Yb}^+$ using the isotopes’ nuclear dipole moments $\mu_{171}=0.4923(4)\mu_N$ and $\mu_{173}=-0.6780(6)\mu_N$  \cite{stoneTableRecommendedNuclear2019}, while the $B$-coefficient is taken from theoretical predictions, $B = -4700 \pm 500$ MHz \cite{dzubaHyperfineInducedElectricDipole2016, xiaoHyperfineStructureYb2020}.

To increase the scanning speed, we use the rapid adiabatic passage (RAP) technique \cite{furstCoherentExcitationHighly2020b}.
A RAP sweep across resonance enables robust state excitation even with fast decoherence, as detailed in the Appendix.
With this method, we first identified 15 previously unobserved transition frequencies with a few kHz accuracy.
The $F$ and $m$ quantum numbers of the HF states are inferred from their Zeeman sensitivities.

\begin{table}[]
    \caption{\label{table:abs_freq}
    Measured HF splittings of $^2F_{7/2}$ state referenced to its $F_e=6$ substate ($W_{F_e}-W_6$). 
    The $F_e=2,3,5,6$ states are interrogated by coherent excitation of the first order magnetic insensitive $\Delta m_F=0$ transitions.
    By exciting the magnetic sensitive transitions using the RAP technique, the uncertainties due to quadratic Zeeman shift of the $F_e=1,4$ states are reduced.
    In addition, we report the absolute \ybd~transition frequencies $\nu^{(F_e)}$ from the $|^2S_{1/2}, F_g=3\rangle$ ground state to the $|^2F_{7/2}, F_e=2,6\rangle$ excited states from an independent measurement compared to the measured \ybz~clock transition.
    The statistical ($u_A$) and systematic ($u_B$) uncertainties are shown in brackets, respectively.}
        \begin{ruledtabular}
            \begin{tabular}{lll}
                \textrm{$F_e$}&  \textrm{$W_{F_e}-W_6$ (Hz)}  & \textrm{$\nu^{(F_e)}-642.1~\mathrm{THz}$ (Hz)} 
                \\
                \colrule
                1 & 3 659 258 477 (71) (503) \\
                2 & 4 487 804 282 (2) (67)  & 23 664 367 825 (16) (435) \\
                3 & 5 232 926 857 (4) (5290) & \\
                4 & 5 297 389 527 (4) (5430) & \\
                5 & 3 884 874 642 (2) (175) \\
                6 & 0 & 19 176 563 542 (16) (432) \\

            \end{tabular}
        \end{ruledtabular}
\end{table}

To determine the absolute frequencies of the clock transitions at the kHz level, we use the E3 transition in \ybz~as a reference (the level scheme can be found in Appendix), whose absolute frequency is known with an uncertainty of 11 Hz, with $\nu_{172}=$ 642 116 785 150 892.3 (11.1) Hz \cite{furstCoherentExcitationHighly2020b}. In this work, we measured a magnetic field drift of $0.07~\mathrm{\mu T}$ over one day, corresponding to a measurement uncertainty of $\sigma^{(172)}_\mathrm{total} = 432$ Hz. This represents the main contribution to the systematic uncertainty budget, shown in Appendix.

For the absolute frequency measurements, we prepare the atomic population in the $|^2S_{1/2}, F_g = 3, m_g = 0\rangle$ ground state and coherently excite transitions to the $|^2F_{7/2}, F_e = 2, 4, 6, m_e = 0\rangle$ states. Details of the experimental sequence are provided in Appendix. A summary of the measured transition frequencies and hyperfine splittings is provided in Table~\ref{table:abs_freq}, and the leading systematic shifts of the absolute frequencies are listed in Appendix.
To determine additional HF splittings and reduce systematic uncertainty, we also prepare the population in the $|^2S_{1/2}, F_g = 2, m_g = 0\rangle$ or $|^2S_{1/2}, F_g = 3, m_g = \pm3\rangle$ states, enabling interrogation of further transitions.
A detailed overview of the measurement results can be found in Appendix. The derivation of the HF coefficients from the measured HF splittings is beyond the scope of this letter and will be discussed in a separate paper \cite{prakashBePublished}, which involves novel atomic structure theory.
 
Due to the large nuclear electric quadrupole moment of $^{173}\mathrm{Yb}$, the HF states of the $^2F_{7/2}$ clock state and $^2S_{1/2}$ ground state mix with intermediate states with the same hyperfine $F$ number, which creates additional electric dipole-allowed decay (HFE1, details in Appendix channels to the ground state. 
In particular, the mixing between the $^2F_{7/2}$ and $^2P_{3/2}$ states is expected to contribute most significantly to the HFE1 channel due to their small energy separation. The decay rate of an excited HF state, $\mathcal{R}^{(F_e)}$, given by the sum of the E3 and HFE1 transition rates ($\mathcal{R}^{(F_e\rightarrow F_g)}_\mathrm{E3}$ and $\mathcal{R}^{(F_e\rightarrow F_g)}_\mathrm{HFE1}$) as detailed in Appendix, can thus be significantly enhanced. To quantify this enhancement, we measure the laser-induced oscillation frequencies (Rabi frequencies $\Omega^{(F_g\rightarrow F_e)}$) between the ground and excited states for several hyperfine levels.

In a trapped-ion system, accurately determining the Rabi frequency from experimental data requires accounting for decoherence effects such as thermal dephasing, magnetic field fluctuations, and laser phase noise. As detailed in Appendix, thermal dephasing is modeled using the Jaynes-Cummings (JC) model to describe the excitation probability. It arises from the thermal distribution over motional quantum states, characterized by the mean phonon number $\bar{n}$, which leads to different Rabi frequencies and results in damping of the observed Rabi oscillations. Magnetic fluctuations and laser phase noise decoherence are modeled with an added exponential decay term involving the decoherence time $t_{\text{decoh}}$. The Rabi frequency is then extracted by fitting the experimental data using this combined model, as described in Appendix.

\begin{figure}[]
	\centering\includegraphics[width=0.45\textwidth]{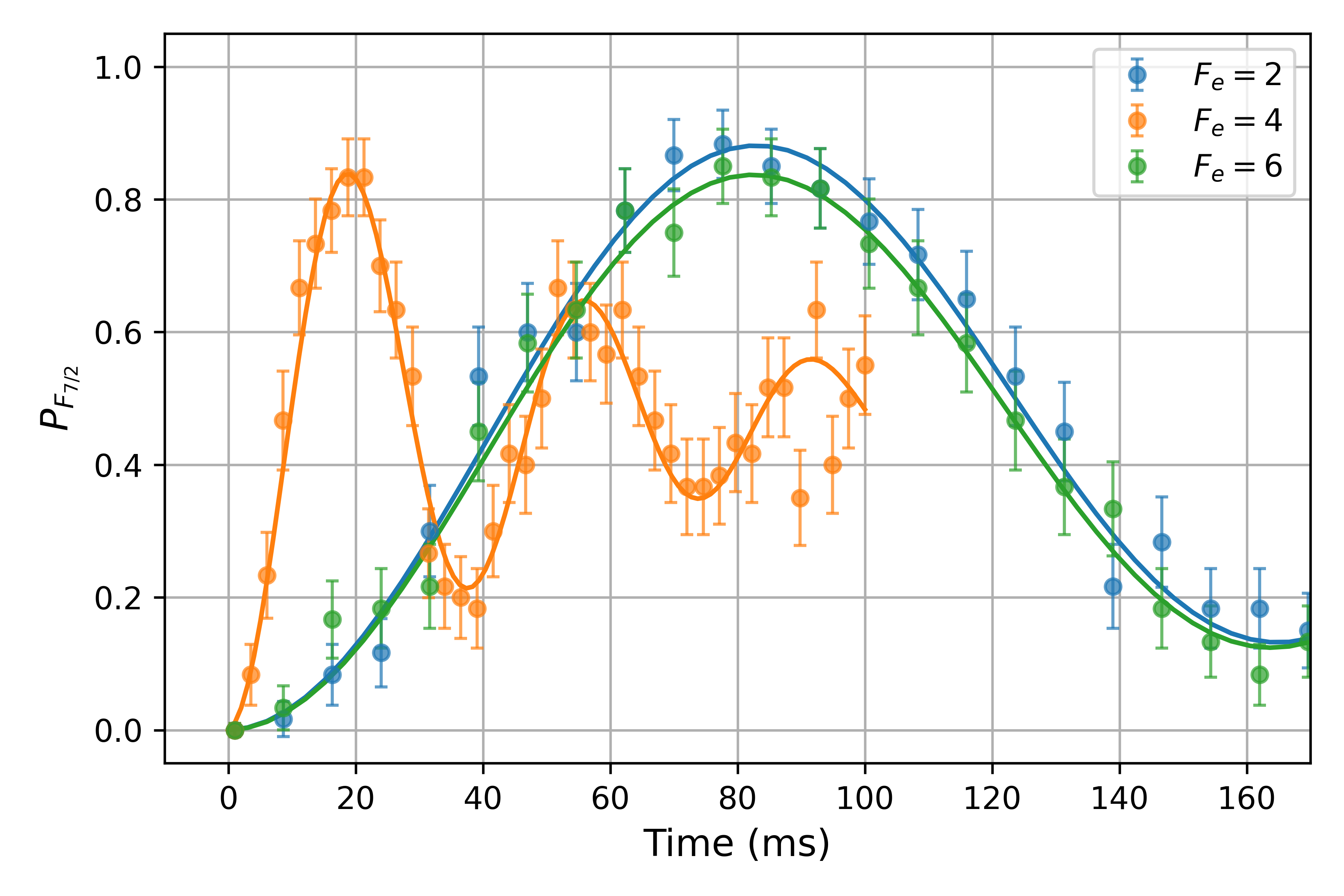}
	\caption{\label{fig:rabi_flop}
    Experimentally measured Rabi oscillations for clock transitions from  $|^2S_{1/2}, F_g=3, m_g=0 \rangle$ ground state to $^2F_{7/2}$ $F_e=2$ (blue), $F_e=4$ (orange) and $F_e=6$ (green) HF states. The experimental data (circle) is fitted with the Jaynes-Cumming model (solid line) with $\bar{n} = 17.6(1.4)$, leading to $\Omega^{(3\rightarrow2)}~=~2\pi~\times~6.74(5)$ Hz, $\Omega^{(3\rightarrow4)}= 2\pi~\times~30.12(36)$~Hz, and $\Omega^{(3\rightarrow6)}= 2\pi\times~6.78(5)$ Hz. The error bar shows the QPN. The quenched transition ($F_e=4$) can be driven at a much higher $\Omega$ with the same optical intensity.}
\end{figure}

The measurement was performed at a well-controlled ion temperature, achieved by carefully adjusting the cooling beam power and by maintaining stable environmental conditions, enabling reliable extraction of both the decoherence time $\tau_{\mathrm{decoh}}$ and the mean phonon number $\bar{n}$. We coherently drive the first-order magnetically insensitive transitions using a 467 nm clock laser with polarization parallel to the magnetic field (see Appendix).
The $|^2S_{1/2}, F_g=3\rangle \rightarrow |^2F_{7/2}, F_e\rangle$ transitions were interrogated in the order $F_e = 4 \rightarrow 2 \rightarrow 6 \rightarrow 4$.
An ion temperature of 0.56(4) mK, close to the Doppler limit and corresponding to a mean phonon number $\bar{n} = 17.6(1.4)$, was confirmed by fitting the Rabi oscillations of the transitions to $F_e = 2$ and $F_e = 6$ using the JC model.
Representative Rabi oscillations for the interrogated transitions are shown in Fig.~\ref{fig:rabi_flop}.
From these measurements, we extracted decoherence times of $\tau_{\mathrm{decoh}} = 1777(865)$ ms, 108(22) ms, and 2370(1425) ms for the transitions to the $F_e =$ 2,4, and 6 hyperfine states, respectively.
The observed decoherence is most likely caused by residual magnetic field fluctuations and the differing quadratic Zeeman sensitivities of the HF states (see Appendix). The uncertainty in the absolute laser intensity cancels in the relative measurement; only intensity fluctuations due to beam pointing at the time scale between two measurements contribute to the final uncertainty. The extracted values of $\Omega^{(3\rightarrow4)}$ from the first and last measurements, 17.2(2) ms and 17.3(2) ms respectively, indicate that laser intensity variation due to beam pointing contributes less than 0.8\% to the overall uncertainty. The validity of the model is proved by a repeated measurement at an increased ion temperature, which shows consistent results, as detailed in Appendix.

After measuring the Rabi frequencies, we relate them to the corresponding transition rates.
The $|^2S_{1/2}, F_g~=~3\rangle \rightarrow |^2F_{7/2}, F_e = 6\rangle$ transition is an unquenched pure E3 transition and serves as a reference. Its lifetime is taken as 1.6(1) years, as previously measured in \ybe \cite{langeLifetime722021,szSurzhykovPrivComm}.
From this lifetime, we extract the transition rate $\mathcal{R}^{(F_e\rightarrow 3)}_\mathrm{E3}$ for each $F_e$ using angular momentum reduction (see Appendix). Results are listed in Table~\ref{table:transition_rates}.  

If the $F_g = 3 \rightarrow F_e = 2,4$ transitions had only E3 contributions, these transition rates $\mathcal{R}^{(F_e\rightarrow 3)}_\mathrm{E3}$ can be used to calculate their expected Rabi frequencies $\Omega^{(3 \rightarrow F_e)}_\mathrm{E3}$ (see Appendix), yielding predicted ratios of $\Omega^{(3 \rightarrow F_e)}_\mathrm{E3}/\Omega^{(3 \rightarrow 6)} = 0.25$ for $F_e = 2$ and 0.51 for $F_e = 4$. 


\begin{table}[]
\caption{\label{table:transition_rates}
Comparison of experimental and theoretical quantities for $F_e=2,4$.
Measured Rabi frequency ratios $\Omega^{(3  \rightarrow F_e)}/\Omega^{(3 \rightarrow 6)}$ are used to derive effective E1 rates $\mathcal{R}^{(F_e\rightarrow 3)}_\mathrm{E1,eff}$ via Eq.~\ref{eq:omegaHFE1} and Eq.~\ref{eq:omegaE3}, which are compared to E3 rates $\mathcal{R}^{(F_e\rightarrow 3)}_\mathrm{E3}$ from the measured lifetime~\cite{langeLifetime722021,szSurzhykovPrivComm}, and to theoretical HFE1 rates $\mathcal{R}^{(F_e\rightarrow 3)}_\mathrm{HFE1}$ from Ref.~\cite{dzubaHyperfineInducedElectricDipole2016}. Uncertainties include Rabi frequency measurements and the E3 lifetime \cite{langeLifetime722021,szSurzhykovPrivComm}. Rates are in $10^{-8}$/s.}
\begin{ruledtabular}
\begin{tabular}{lcccc}
$F_e$ & $\Omega^{(3 \rightarrow F_e)}/\Omega^{(3 \rightarrow 6)}$ & $\mathcal{R}^{(F_e\rightarrow 3)}_\mathrm{E1,eff}$ & $\mathcal{R}^{(F_e\rightarrow 3)}_\mathrm{E3}$ & $\mathcal{R}^{(F_e\rightarrow 3)}_\mathrm{HFE1}$ \\
\hline
2 &  0.987(35)  & $1.73(16)$   &   0.29(1)    & 51   \\
4 &  4.22(20)  & $23.7(3.0)$   &   0.96(5)    & 650   \\
\end{tabular}
\end{ruledtabular}
\end{table}

The experimentally measured ratios $\Omega^{(3 \rightarrow F_e)}/\Omega^{(3 \rightarrow 6)}$, listed in Table \ref{table:transition_rates}, are significantly larger, confirming that the HFE1 decay channel is dominating in these transitions, as discussed in Appendix.
Hence we assume that the E3 contribution is small compared to the HFE1 contribution for $F_e = 2,4$, and treat these as effectively pure E1 transitions, as detailed in Appendix. Under this approximation, the effective E1 rates $\mathcal{R}^{(F_e \rightarrow 3)}_\mathrm{E1,eff}$ are derived from the measured ratio between Rabi frequencies $\Omega^{(3 \rightarrow F_e)}/\Omega^{(3 \rightarrow 6)}$ by combining Eq.~\ref{eq:omegaHFE1}-\ref{eq:rate_E3} (see Appendix).
The inferred effective rate for the $F_g = 3 \rightarrow F_e = 2$ transition is approximately 6 times larger than the E3 transition rate, and about 25 times larger for $F_e=4$, indicating a strong enhancement of the $F_e = 2$ and $F_e = 4$ decay rates due to the HFE1 contribution. As discussed in Appendix, this enhancement leads to the intriguing effect, that the excited HF states have individual lifetime, while identical lifetime is expected for unperturbed hyperfine states. 
By neglecting the E3 contribution, the inferred lifetime of the $F_e = 4$ state is about 12 times shorter than the unquenched state. Since $F_e = 4 \rightarrow F_g=2$ is not quenched, the lifetime of this state is approximately $1/\mathcal{R}^{(4\rightarrow3)}_\mathrm{eff,E1} = 49(21)$ days, scaled from the 1.6(1)-year lifetime of \ybe \cite{langeLifetime722021,szSurzhykovPrivComm} (see Appendix).

Additionally, we compare the effective rates with the theoretically estimated HFE1 rates $\mathcal{R}^{(F_e \rightarrow 3)}_\mathrm{HFE1}$ from \cite{dzubaHyperfineInducedElectricDipole2016}, as summarized in Table~\ref{table:transition_rates}. 
The experimentally inferred effective transition rates differ from the theoretical values by a factor of about 28. 
Checks were also performed as summarized in Appendix.
The error introduced by approximating $\mathcal{R}^{(F_e \rightarrow 3)}_\mathrm{E1,eff}$ as $\mathcal{R}^{(F_e \rightarrow 3)}_\mathrm{HFE1}$ is, in our case and based on Table~\ref{table:transition_rates}, at most 50\% (see Appendix). A potential change in the inferred lifetime may arise from the taken reference \cite{langeLifetime722021,szSurzhykovPrivComm}, i.e., the experimental lifetime of the $^2F_{7/2}$ clock state of \ybe, which is currently under re-evaluation \cite{szSurzhykovPrivComm}.

\begin{figure}[]
	\centering\includegraphics[width=0.48\textwidth]{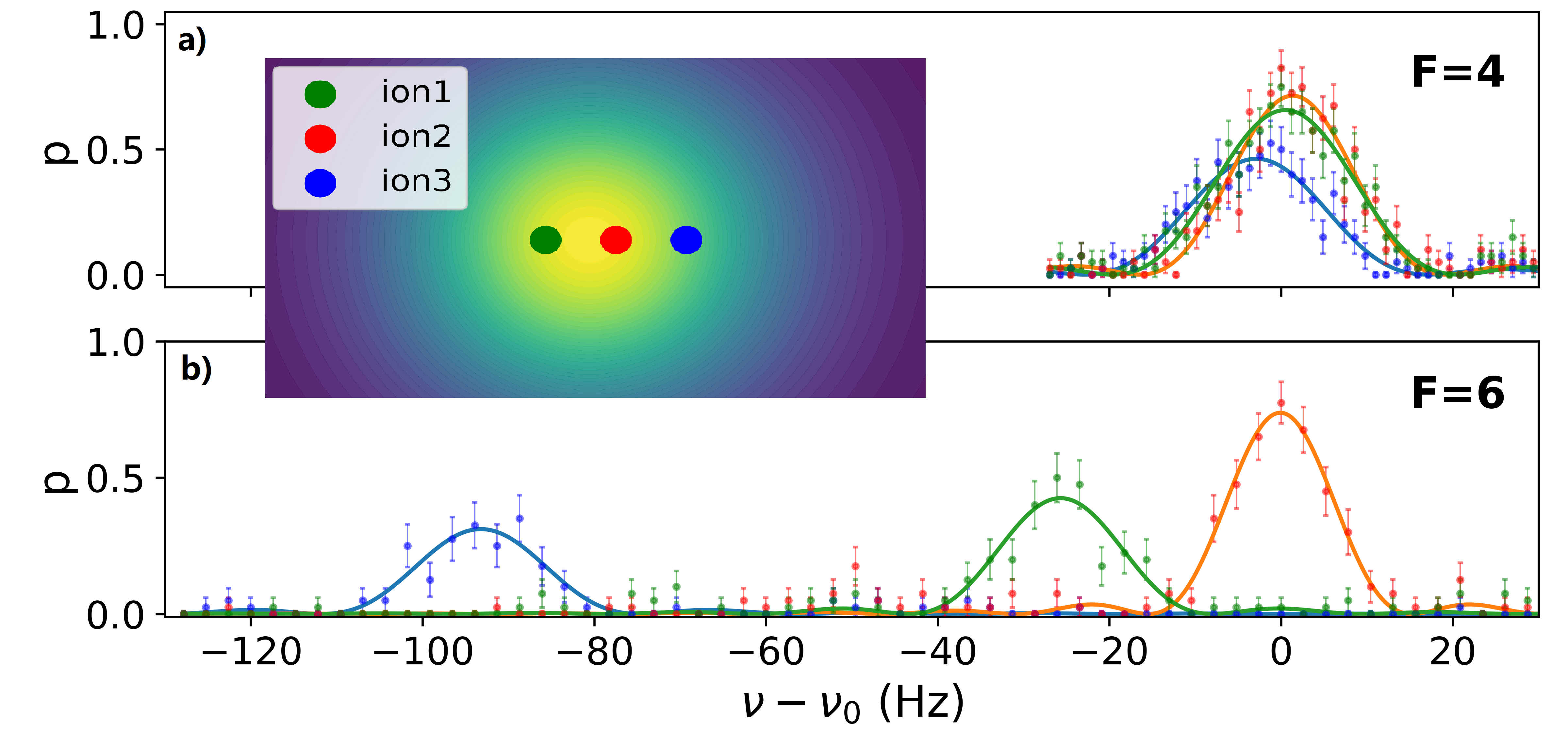}
	\caption{\label{fig:freq_scan}
    Experimentally measured clock resonances of a 3-ion Coulomb crystal with $10~\mathrm{\mu m}$ inter-ion spacing. The inset illustrates the laser intensity and ion positions (not to scale).
    The laser beam radius is $w = 30~\mathrm{\mu m}$, pulse time $\tau=55$ ms and  $\Omega^{(3\rightarrow6)}\approx\Omega^{(3\rightarrow4)}= 2\pi \times 9.1(1.2)$ Hz. The relative intensity between the two measurements is $I^{(3\rightarrow6)}_\mathrm{Laser}/I^{(3\rightarrow4)}_\mathrm{Laser}=20.6(1.5)$.
    Referencing to the transition frequency of ion 2, $\nu_\mathrm{ion2}^{(F_g\rightarrow F_e)}$, we observe a  differential AC Stark shift of $\nu_\mathrm{ion1}^{(3\rightarrow4)}=-0.9(4)$ Hz, $\nu_\mathrm{ion3}^{(3\rightarrow4)}=-4.2(4)$~Hz, $\nu_\mathrm{ion1}^{(3\rightarrow6)}=-25.8(4)$ Hz and $\nu_\mathrm{ion3}^{(3\rightarrow6)}=-93.3(4)$ Hz.
    Hence for a similarly Fourier-limited linewidth $\Delta\nu\approx16$~Hz, the transition from $F_g=3$ to $F_e=4$ state (Fig.~3a)  requires less laser intensity than to the $F_e=6$ state (Fig.~3b), which results in a smaller AC Stark shift. }
\end{figure}

To demonstrate the advantage of the quenched transition ($F_g=3\rightarrow F_e = 4$), we trap a 3-ion Coulomb crystal and compare spectra to the reference transition ($F_g=3 \rightarrow F_e = 6$).
Both transitions are driven by a Gaussian beam, with identical pulse durations but different probe beam powers chosen to achieve the same Rabi frequency.
We observe that for the same Rabi frequency, the quenched transition requires significantly less optical intensity than the reference transition, resulting in smaller AC Stark shifts on the transition frequency.
As visualized in Fig.~\ref{fig:freq_scan}a, multi-ion operation is feasible with the quenched transition, despite of the large intensity gradient of the Gaussian beam.
In contrast, simultaneous excitation of multiple ions is not possible with the reference transition (Fig.~\ref{fig:freq_scan}b) due to the larger differential AC Stark shift.
In this measurement, the extracted reduction factor of the AC Stark shift is 22(5), in good agreement with the value extrapolated from the measured Rabi frequency ratio, 18(2).

As a candidate for future optical clocks, the quadratic Zeeman shift may become a limiting factor for accuracy.
Therefore, based on the measured hyperfine splittings of the $^2F_{7/2}$ state in Table \ref{table:abs_freq}, we calculate the second-order Zeeman sensitivities.
Among the first-order magnetically insensitive transitions, the $F_g=3 \rightarrow F_e = 6$ transition exhibits the smallest second-order Zeeman sensitivity of $-33.7~\mathrm{mHz/\mu T^2}$.
Notably, the small energy gap between the $F_e = 3$ and $F_e = 4$ hyperfine states leads to large second-order Zeeman sensitivities of $-7012~\mathrm{mHz/\mu T^2}$ and $8060~\mathrm{mHz/\mu T^2}$, respectively.
These sensitivities, much larger than the value of $-2.03(2)~\mathrm{mHz/\mu T^2}$ reported for \ybe $F_g=0 \rightarrow F_e = 3$ transition \cite{huntemannHighaccuracyOpticalClock2014a}, may limit the interrogation time during clock operation due to magnetic decoherence.
Such decoherence can be suppressed by operating the clock at a lower magnetic field while interrogating the transition.
The first and second order Zeeman sensitivities of all hyperfine and Zeeman substates are summarized in Appendix. 
  
In summary, we measured the hyperfine structure of the $^2F_{7/2}$ state in \ybd, from which we calculated its second-order Zeeman sensitivities.
The transition frequency of $|^2S_{1/2}, F_g=3\rangle \rightarrow |^2F_{7/2}, F_e=6\rangle$ (unquenched) is determined to be $642.11917656354(43)$~THz.
We also calculated the HFE1 transition rates using approximations detailed in Appendix, showing that for the $F_g=3\rightarrow F_e = 4$ transition, the inferred HFE1 rate is significantly higher than the E3 rate. This leads to a 12 times shorter lifetime for the $F_e=4$ hyperfine state, revealing hyperfine-dependent, nuclear spin–induced quenching of this clock transition. Additionally, we demonstrate simultaneous excitation of multiple ions using a Gaussian beam, highlighting the potential of \ybd for future multi-ion optical clocks.

Compared to the E3 transition of \ybe, the $|^2S_{1/2}, F_g=3 \rangle \rightarrow |^2F_{7/2}, F_e=4\rangle$ transition of \ybd with a lifetime of tens of days, can be driven at a 3.67(17) higher Rabi frequency. Therefore, this quenched transition could serve as a interesting shelving transition for future quantum computers. 
Also, the AC Stark shift at given Rabi frequency is reduced by more than one order of magnitude.
This would enable simultaneous interrogation of up to 100 ions using a segmented Paul trap as described in Ref. \cite{herschbachLinearPaulTrap2012b, kellerProbingTimeDilation2019a} and a flat-top beam \cite{yuPrecisionSpectroscopyYb2024a}.
The large number of ions would facilitate the implementation of advanced clock protocols, such as zero-dead time clocks \cite{biedermannZeroDeadTimeOperationInterleaved2013a, schioppoUltrastableOpticalClock2017} or cascaded clocks \cite{borregaardEfficientAtomicClocks2013b, humeProbingLaserCoherence2016a} for transportable optical clock systems.
Such clocks could play a pivotal role in intercontinental clock comparison, contributing to the future redefinition of the SI second \cite{dimarcqRoadmapRedefinitionSecond2024} and to relativistic geodesy \cite{grottiLongdistanceChronometricLeveling2024, takamotoTestGeneralRelativity2020a,mehlstaublerAtomicClocksGeodesy2018a}.
     
However, the accuracy and the interrogation time of multi-ion optical clocks based on the clock transition $|^2S_{1/2}, F_g=3\rangle\rightarrow|^2F_{7/2}, F_e=4\rangle$ are likely to be limited by the strong quadratic Zeeman shift.
Employing HF-averaging technique \cite{barrettDevelopingFieldIndependent2015b}, operation under field-insensitive conditions \cite{arnoldSuppressionClockShifts2016a} or cascaded clock protocols combining high stability and high accuracy clock transitions \cite{Kim2023} can improve both the accuracy and the coherence time of such a multi-ion optical clock.

\begin{acknowledgments}
We kindly acknowledge Thomas Legero, Erik Benkler and Nils Huntemann for providing ultra-stable laser light at 934 nm.
We thank Christian Sanner, Wesley C. Campbell, Julian Berengut, Jacinda Ginges, Andrey Surzhykov, Jonas Keller and Ingrid Richter for helpful discussions.
This project has been supported by the Deutsche Forschungsgemeinschaft (DFG, German Research Foundation) through grant CRC SFB 1227 (DQ-mat, project B03) and through Germany's Excellence Strategy EXC-2123 QuantumFrontiers - 390837967.
We acknowledge support by the project 22IEM01 TOCK, which has received funding from the European Partnership on Metrology, co-financed from the European Union’s Horizon Europe Research and Innovation Programme and by the Participating States.
This work is also supported by Funding support from the NSRF via the Program Management Unit for Human Resources \& Institutional Development, Research and Innovation [grant number B39G680007] and by the Max-Planck-RIKEN-PTB-Center.
 
\end{acknowledgments}%

%

\clearpage


\appendix

\section{Experimental geometry}

\begin{figure}[h]
	\centering\includegraphics[width=0.35\textwidth]{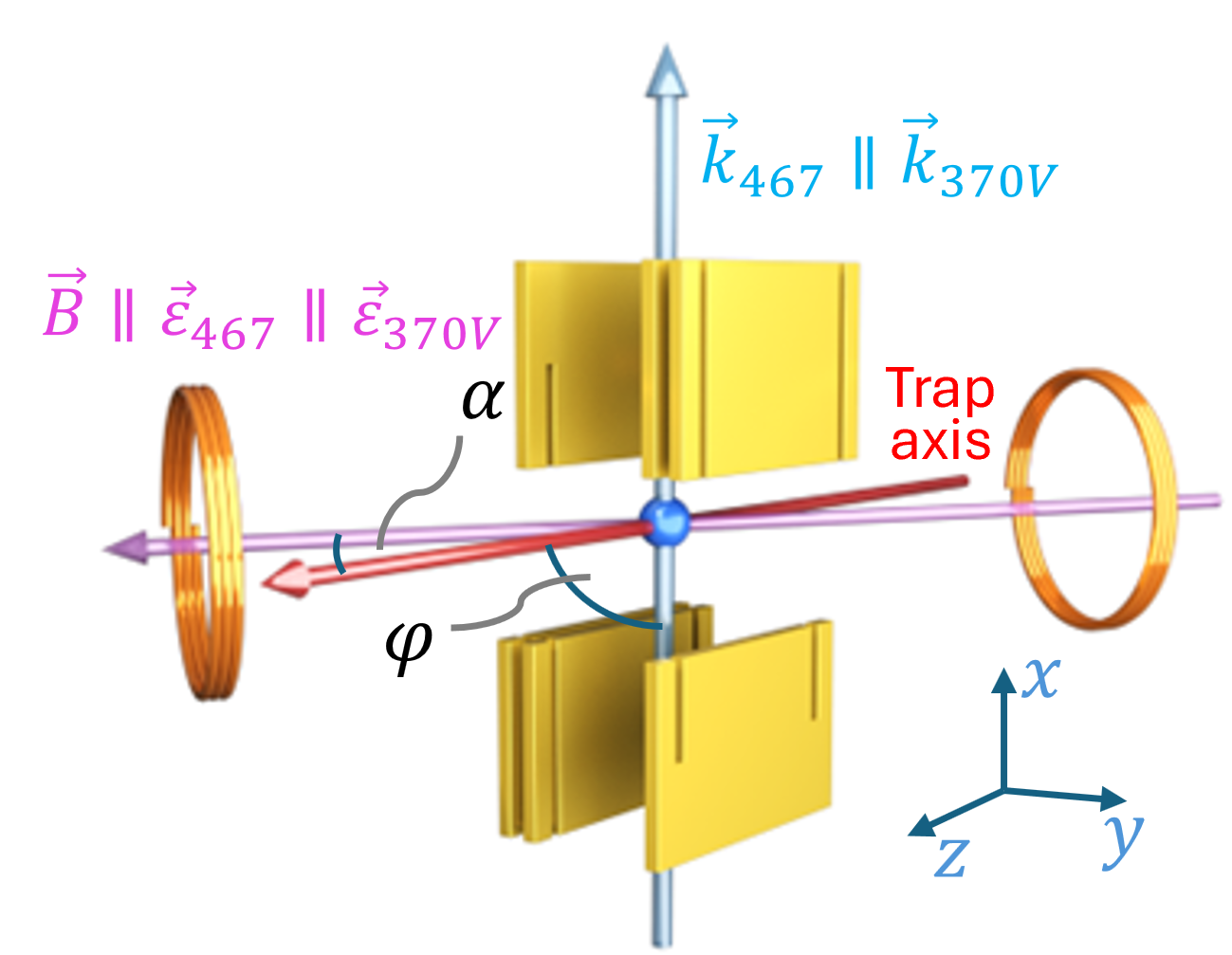}
	\caption{\label{fig:trap}
		Experimental setup. A $^{173}\mathrm{Yb}^{+}$ ion (blue sphere) is trapped in a linear Paul trap. The red arrow indicates the axial direction of the RF trap, which coincides with the $z$-axis. The blue arrow shows the propagation direction of the vertical 370 nm beam (370V) and the 467 nm beam. The polarization of both the 370V and 467 nm beams is aligned with the magnetic field (pink arrow). The angle between the spectroscopy beam and the trap axis, $\varphi$, is 90$^{\circ}$, and the angle between the magnetic field and the trap axis, $\alpha$, is approximately 26$^{\circ}$.}
\end{figure}
The geometry of the experimental setup for laser spectroscopy on trapped \ybd~ions is shown in Fig.~\ref{fig:trap}. The quantization axis is defined by the applied magnetic field at an angle of 26$^{\circ}$ to the $z$-axis.
In addition to the laser beams depicted in Fig.~\ref{fig:trap}, 935~nm and 760~nm repumping lasers are sent along the trap axis (i.e., the $z$-axis). A linearly polarized 370 nm beam, aligned with the magnetic field direction, assists Doppler cooling.
A 411 nm laser beam propagating along the $x$-axis can be used for probing the $|^2S_{1/2}\rangle \rightarrow |^2D_{5/2}\rangle$ transition, with its polarization vector $\vec{\epsilon}_{411}$ perpendicular to the magnetic field.

\section{Transition Rates, Decay Rates, and Rabi Frequency}

In both the main text and in Appendix, we use the terms Rabi frequency, transition rate, and decay rate. Here we clarify their distinctions.

\begin{figure}[h]
\centering\includegraphics[width=0.35\textwidth]{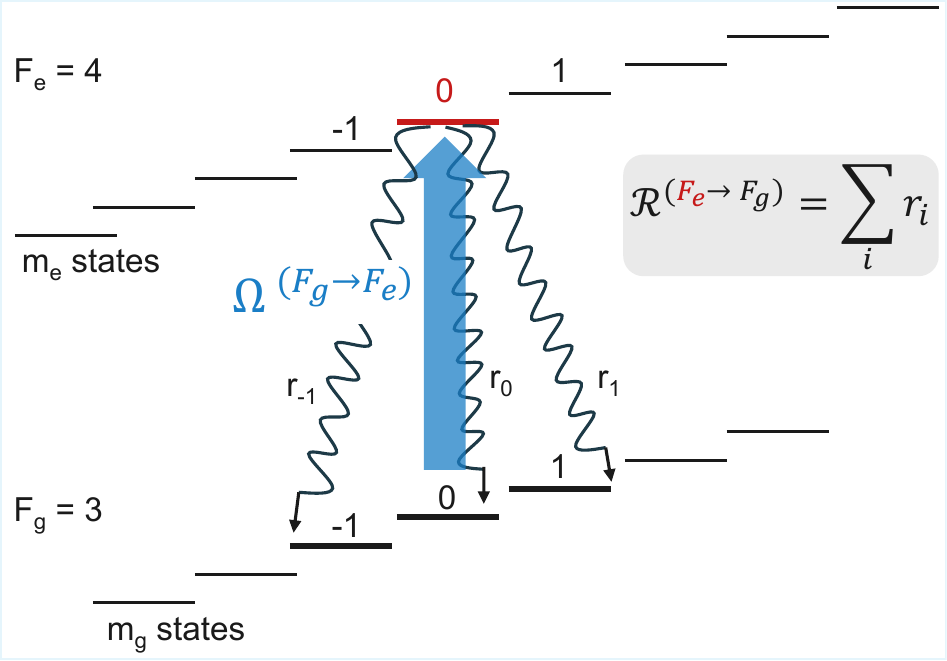}
\caption{\label{fig:rabi_frequency_cropped}
Illustration of the Rabi frequency and transition rate for the $F_e=4, m_e=0 \rightarrow F_g=3$ case, with an E1 decay channel. Each Zeeman sublevel $m_e$ of the excited state has the same transition rate $\mathcal{R}^{(F_e \rightarrow F_g)}$, obtained by summing over the rates of all allowed decay channels to the ground state (here $r_0 + r_{1} + r_{-1}$).}
\end{figure}

The \textbf{transition rate} $\mathcal{R}^{(F_e \rightarrow F_g)}$ between hyperfine states $F_e$ and $F_g$ is defined as the total rate from the excited state to all allowed Zeeman sublevels of $F_g$.
The \textbf{Rabi frequency} $\Omega^{(F_g \rightarrow F_e)}$ characterizes the coherent excitation strength of a specific driven transition between individual Zeeman sublevels of $F_g$ and $F_e$.
The \textbf{decay rate} $\mathcal{R}^{(F_e)}$ of an excited state $F_e$ is the sum of all allowed transition rates to ground states

\begin{equation}
    \mathcal{R}^{(F_e)} = \sum_{F_g} \mathcal{R}^{(F_e \rightarrow F_g)} = \frac{1}{\tau^{(F_e)}},
\end{equation}
and is the inverse of the excited-state lifetime $\tau^{(F_e)}$.
The relation between the Rabi frequency and the transition rate is discussed below.

\subsection{Hyperfine-induced electric dipole (HFE1)}

We treat the hyperfine-induced electric dipole (HFE1) contribution as an effective E1 process. Since we use $\pi$-polarized light, the Rabi frequency can be written as
\begin{equation}
\Omega^{(F_g \rightarrow F_e)}_{\text{HFE1}} = \frac{eE_0}{\hbar} \,
\frac{\langle F_g m_g, 1,0 \mid F_e m_e\rangle}{\sqrt{2F_e+1}}
\langle F_e \,\|\, \hat{r} \,\|\, F_g\rangle,
\label{eq:omegaHFE1}
\end{equation}
where $e$ is the elementary charge, $E_0$ is the electric field amplitude of the clock laser, $\langle F_g m_g, 1,0 \mid F_e m_e\rangle$ and $\langle F_e \,\|\, \hat{r} \,\|\, F_g\rangle$ are the Clebsch-Gordan coefficient and the reduced transition matrix element, respectively.
The corresponding HFE1 transition rate is \cite{dzubaHyperfineInducedElectricDipole2016}
\begin{equation}
    \mathcal{R}^{(F_e \rightarrow F_g)}_{\text{HFE1}} = \frac{4}{3} \,
c \alpha k^3 \frac{\mid\langle F_e \,\|\, \hat{r} \,\|\, F_g\rangle\mid^2}{2F_e + 1},
\label{eq:rate_HFE1}
\end{equation}
where $c$ is the speed of light, $\alpha$ the fine-structure constant, and $k$ is the wavevector of the clock laser. Eq. \ref{eq:omegaHFE1} and Eq. \ref{eq:rate_HFE1} show that the Rabi frequency is proportional to $\sqrt{\mathcal{R}^{(F_e \rightarrow F_g)}_{\text{HFE1}}}$. In the main text, we use measured Rabi frequencies $\Omega^{(3 \rightarrow F_e)}$ for $F_e = 2,4$ to calculate effective E1 transition rates $\mathcal{R}_{\mathrm{E1,eff}}^{(F_e\rightarrow 3)}$ using Eq.~\ref{eq:omegaHFE1} and \ref{eq:rate_HFE1}, under the hypothesis that $\mathcal{R}_{\mathrm{HFE1}}^{(F_e\rightarrow F_g)} \approx \mathcal{R}_{\mathrm{E1,eff}}^{(F_e\rightarrow F_g)}$.

\subsection{Electric octupole (E3)}

For E3 transitions between $m_g=0$ and $m_e=0$ Zeeman substates, excited with $\pi$-polarized light the Rabi frequency is
\begin{equation}
\Omega^{(F_g \rightarrow F_e)}_{\text{E3}} = \frac{eE_0}{\hbar} \frac{k^2}{30} \,
\frac{\langle F_g , 0, 3,0 \mid F_e, 0\rangle}{\sqrt{2F_e+1}}
\langle F_e \,\|\, \hat{O}(r^3) \,\|\, F_g\rangle,
\label{eq:omegaE3}
\end{equation}
with $\hat{O}(r^3)$ the octupole operator. The additional factor $k^2$ arises from second order spatial derivative of the laser's electric field. The factor 30 is a geometric factor that depends on $m_e-m_g$, the angles betweeen the polarization vector, the wavevector and the quantization axis (see Fig.~\ref{fig:trap} and \cite{Johnson2007}). The corresponding transition rate is
\begin{equation}
    \mathcal{R}^{(F_e \rightarrow F_g)}_{\text{E3}} = \frac{8k^4}{4725} \,
c \alpha k^3 \frac{\mid\langle F_e \,\|\, \hat{O}(r^3) \,\|\, F_g\rangle\mid^2}{2F_e + 1},
\label{eq:rate_E3}
\end{equation}
where the coefficient $8k^4/4725$ follows from multipole expansion \cite{Johnson2007}, leading to a $k^7$-dependence combined with the $k^3$ arising from the state density of the photon modes. In order to relate it to the lifetime, we use angular momentum reduction \cite{kingAngularMomentumCoupling2008a}
\begin{equation}
\begin{split}
   \mid\langle F_e \,\|\, \hat{O}(r^3) \,\|\, F_g\rangle\mid
   &= \sqrt{(2F_e+1)(2F_g+1)} \\
   &\quad\times
   \begin{Bmatrix}
      J_e & 3 & J_g \\
      F_g & I & F_e
   \end{Bmatrix}
   \mid\langle J_e \,\|\, \hat{O}(r^3) \,\|\, J_g\rangle\mid,
\end{split}
\label{eq:ang_red}
\end{equation}
and $\{\}$ denotes a 6-$j$ Wigner symbol. 

Finally, the total E3 decay rate is
\begin{equation}
    \mathcal{R}_{\text{E3}} = \frac{1}{\tau_\mathrm{E3}} = \frac{8}{4725} \,
c \alpha k^7 \frac{\mid\langle J_e \,\|\, \hat{O}(r^3) \,\|\, J_g\rangle\mid^2}{2J_e + 1},
\label{eq:decay_rate_E3}
\end{equation}
where $\tau_\mathrm{E3} = 1.6(1)\,\text{years}$ is the measured lifetime of the unperturbed $F_e = 6$ state \cite{langeLifetime722021,szSurzhykovPrivComm}. We combine Eq.~\ref{eq:rate_E3}, \ref{eq:ang_red}, and \ref{eq:decay_rate_E3} to derive the transition rate $\mathcal{R}^{(F_e \rightarrow 3)}_{\text{E3}}$ for $F_e = 2,4$, assuming a pure E3 contribution, which is then given in Table II of the main text.  
We also use Eq.~\ref{eq:omegaE3} to evaluate the ratios $\Omega^{(3 \rightarrow F_e)}_{\text{E3}} / \Omega^{(3 \rightarrow 6)}$ for $F_e = 2$ and $F_e =4$.

\section{Additional measurements}

Since we observed a large discrepancy between the the theoretically predicted HFE1 transition rates $\mathcal{R}_\mathrm{HFE1}$ \cite{dzubaHyperfineInducedElectricDipole2016} and the experimentally determined effective transition rates $\mathcal{R}_\mathrm{E1,eff}$, we performed three additional sets of measurements, summarized in Table~\ref{table:compare_omega}. 1) We compared two sets of unquenched (pure E3) transitions using $\pi$-polarized light. The measured Rabi frequency ratios agree with the calculated values, confirming the stability of the laser intensity and spectral purity between the two measurements, as well as validating the derived E3 transition equations presented in later section. 
2) By switching the laser polarization $\vec{\epsilon}$, we compared the Rabi frequencies of the $\Delta m_F = 0$ ($\vec{\epsilon} \parallel \vec{B}$) and $\Delta m_F = \pm 1$ ($\vec{\epsilon} \perp \vec{B}$) transitions of the same ${}^{2}F_{7/2}$, $F_e = 4$ clock transition. For $\Delta m_F = \pm 1$, the E3 contribution is as low as 2\%  of the measured Rabi frequency, whereas for $\Delta m_F = 0$  the E3 contributes to 12\%. Measurements of transitions with different E3 contributions confirm the approximation made in the main text of neglecting these contributions.
This measurement validates the derived equations for HFE1 transitions and confirms the polarization purity of the clock laser. 3) We coherently excite the $S_{1/2}, F_g=2 \rightarrow F_{7/2},F_e=3$ transition, and compare it with the $F_e=6$ reference transition. The derived effective E1 transition rate is $\mathcal{R}^{(3\rightarrow 2)}_\mathrm{E1,eff}=6.9(1.5)\times10^{-8}$/s. We include this data and calculate the HFE1 transition rates, which might give hints to the possible origin of the discrepancy between the theory and the experiment.

Besides the effective E1 transition rates, we also calculate the possible HFE1 transition rates (see Table \ref{table:HFE1_transition_rates}). Intriguingly, under the assumption of destructive interference between the HFE1 and E3 channels, the ratios between the derived HFE1 transition rate and the theoretical prediction are are nearly identical for all measured transitions. Further theoretical investigation and an independent verification of the \ybe lifetime are needed to resolve this discrepancy.

\begin{table}[h]
\caption{\label{table:compare_omega} Comparison of Rabi frequency ratios. All measurements were performed using the \ybd isotope, except for the transition marked with an asterisk (*), which was measured on the \ybz isotope (unquenched E3 transition).}
\begin{ruledtabular}
\begin{tabular}{ccrr}
\textrm{Transition (1)} &  \textrm{Transition (2)} & \multicolumn{2}{c}{$\Omega^{(1)}/\Omega^{(2)}$}
\\
\hline
$F_g,m_g\rightarrow F_e,m_e$ &  $F_g,m_g\rightarrow F_e,m_e$ & measured &  calculated 
\\
\hline
$-1/2\rightarrow-1/2$* & $3,0\rightarrow6,0$ & 1.13(8) & 1.15
\\
$2,0\rightarrow5,0$&  $3,0\rightarrow6,0$ & 0.56(5) & 0.54
\\
\hline
$3,0\rightarrow4,0$&  $3,0\rightarrow4,+1$ & 1.66(17) & 1.79
\\
$3,0\rightarrow4,0$&  $3,0\rightarrow4,-1$ & 1.72(17) & 1.79
\\
\hline
$2,0\rightarrow3,0$&  $3,0\rightarrow6, 0$ & 2.23(19) & -

\end{tabular}
\end{ruledtabular}
\end{table}

\begin{table}[h]
\caption{\label{table:HFE1_transition_rates} Comparison between experimental and theoretical transition rates. The ratios between the theoretically predicted HFE1 rates $\mathcal{R}_\mathrm{HFE1}$ \cite{dzubaHyperfineInducedElectricDipole2016} and the effective E1 transition rates $\mathcal{R}_\mathrm{E1,eff}$ are calculated, along with the corresponding ratios assuming either destructive ($\mathcal{R}_\mathrm{E1,dest}$) or constructive ($\mathcal{R}_\mathrm{E1,const}$) interferences between the E3 and HFE1 channels.}
\begin{ruledtabular}
\begin{tabular}{cccc}
\textrm{$F_g,m_g \rightarrow F_e,m_e$} & $\frac{\mathcal{R}_\mathrm{HFE1}}{\mathcal{R}_\mathrm{E1,eff}}$ & $\frac{\mathcal{R}_\mathrm{HFE1}}{\mathcal{R}_\mathrm{E1,const}}$ & $\frac{\mathcal{R}_\mathrm{HFE1}}{\mathcal{R}_\mathrm{E1,dest}}$
\\
\hline

$2,0\rightarrow3,0$&  36.3(7.9) & 80.2(25.4) & 23.5(4.2)
\\
$3,0\rightarrow2,0$& 29.5(2.7) & 52.9(6.0) & 18.8(1.5)
\\
$3,0\rightarrow4,0$& 27.4(3.5) & 35.4(5.0) & 21.8(2.5)

\end{tabular}
\end{ruledtabular}
\end{table}

\section{Search for the E3 transition in \ybd}
\begin{figure}[h]
	\centering\includegraphics[width=0.45\textwidth]{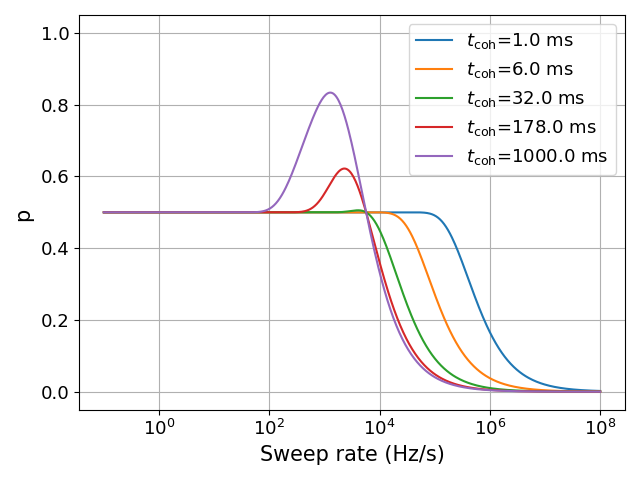}
	\caption{\label{fig:RAP1}
	Simulations of the excitation probability using the RAP technique with a fixed Rabi frequency of $\Omega_{0}~=~2\pi~\times~20$~Hz, shown as colored lines for various coherence times ranging from 1 to 1000 ms.}
\end{figure}

\begin{figure}[h]
	\centering\includegraphics[width=0.45\textwidth]{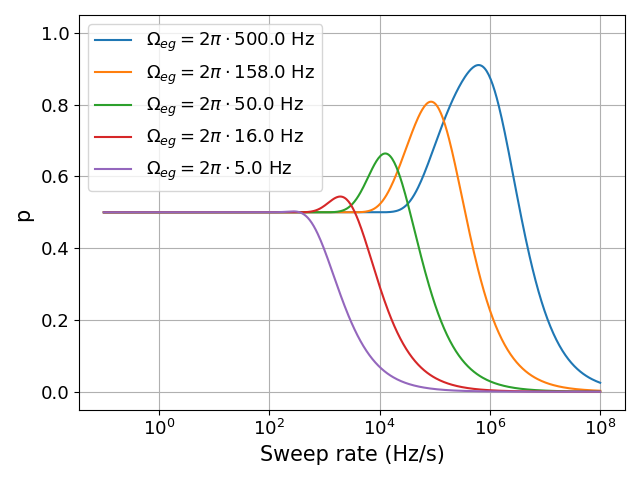}
	\caption{\label{fig:RAP2}
	Simulation of the excitation probability using the RAP technique at a fixed coherence time $t_\mathrm{coh}=100$ ms, with the Rabi frequency varied in the simulation.}
\end{figure}

The rapid adiabatic passage (RAP) technique, previously employed to detect the electric octupole (E3) transition in \ybz ions \cite{furstCoherentExcitationHighly2020b}, is used here to search for the E3 transition in \ybd ions. It enables efficient excitation despite the presence of decoherence.
Because RAP does not rely on resonant excitation, all ions in a Coulomb crystal can be excited simultaneously, despite their different AC Stark shifts.
The excitation probability achieved by RAP can be described using the model \cite{noelAdiabaticPassagePresence2012, furstCoherentExcitationHighly2020b}:
\begin{align}
p=\left(1-e^{-\frac{\Omega_{0} ^2}{4 \alpha }}\right) e^{-\frac{\pi \Omega_{0} }{ \alpha t_{\mathrm{coh}} }}+\frac{1}{2} \left(1-e^{-\frac{\pi \Omega_{0} }{ \alpha t_\mathrm{coh}}}\right)\ ,
\label{eq:rapmodel}
\end{align}
where $\alpha$ is the sweeping rate of the laser frequency and $t_\mathrm{coh}$ is the coherence time. $\Omega_{0} $ is the Rabi frequency as given in Eq.~2 in the main text.
Simulations of different RAP sweeps are shown in Fig. \ref{fig:RAP1} and \ref{fig:RAP2}.


To accelerate the search for the E3 transition, we want to increase the sweep rate while maintaining $p\geq0.5$. 
Numerical simulations (Fig.~\ref{fig:RAP1} and \ref{fig:RAP2}) show that a high excitation probability can be achieved at faster sweep rates by increasing $\Omega_0$ and reducing the coherence time $t_\mathrm{coh}$.
We therefore begin by searching for the transition resonance of the $|^2F_{7/2}, F_e=4, m_e=\pm 1\rangle$ state, for the following reasons: 1) This transition is predicted to have the largest $\Omega_0$.
2) The $m_e=\pm 1$ Zeeman substates have a shorter coherence time.
3) The frequency splitting between $F_e=3$ and $F_e=4$ states is predicted to be small, thus the chance of finding at least one of both transitions in a 100 MHz range is high.
4) The $F_e=3$ state is expected to have a similarly large $\Omega_0$ as the $F_e=4$ state. Although the $|^2S_{1/2},F_g~=~3, m_g~=~0~\rangle~\rightarrow~|^2F_{7/2},F_e~=~3, m_e~=~0\rangle$ transition is forbidden, the $m_g=0 \rightarrow m_e=1$ transition remains allowed from the prepared ground state due to HFE1 selection rules.
For the search, we assume a coherence time of $t_\mathrm{coh} = 200$ ms and choose a sweep rate of $\alpha = 5$ kHz/s over a 2 s-long pulse.
To further enhance the signal-to-noise ratio, we trap a Coulomb crystal of three ions.
The laser intensity introduces an estimated AC Stark shift of about 300 Hz, which can be covered by the wide sweep range of the RAP technique. 
The total spin $F$ of each hyperfine state is inferred from its Zeeman sensitivity (see Table~\ref{table:zeeman}).

\section{Level scheme}

\begin{figure}[h]
	\centering\includegraphics[width=0.45\textwidth]{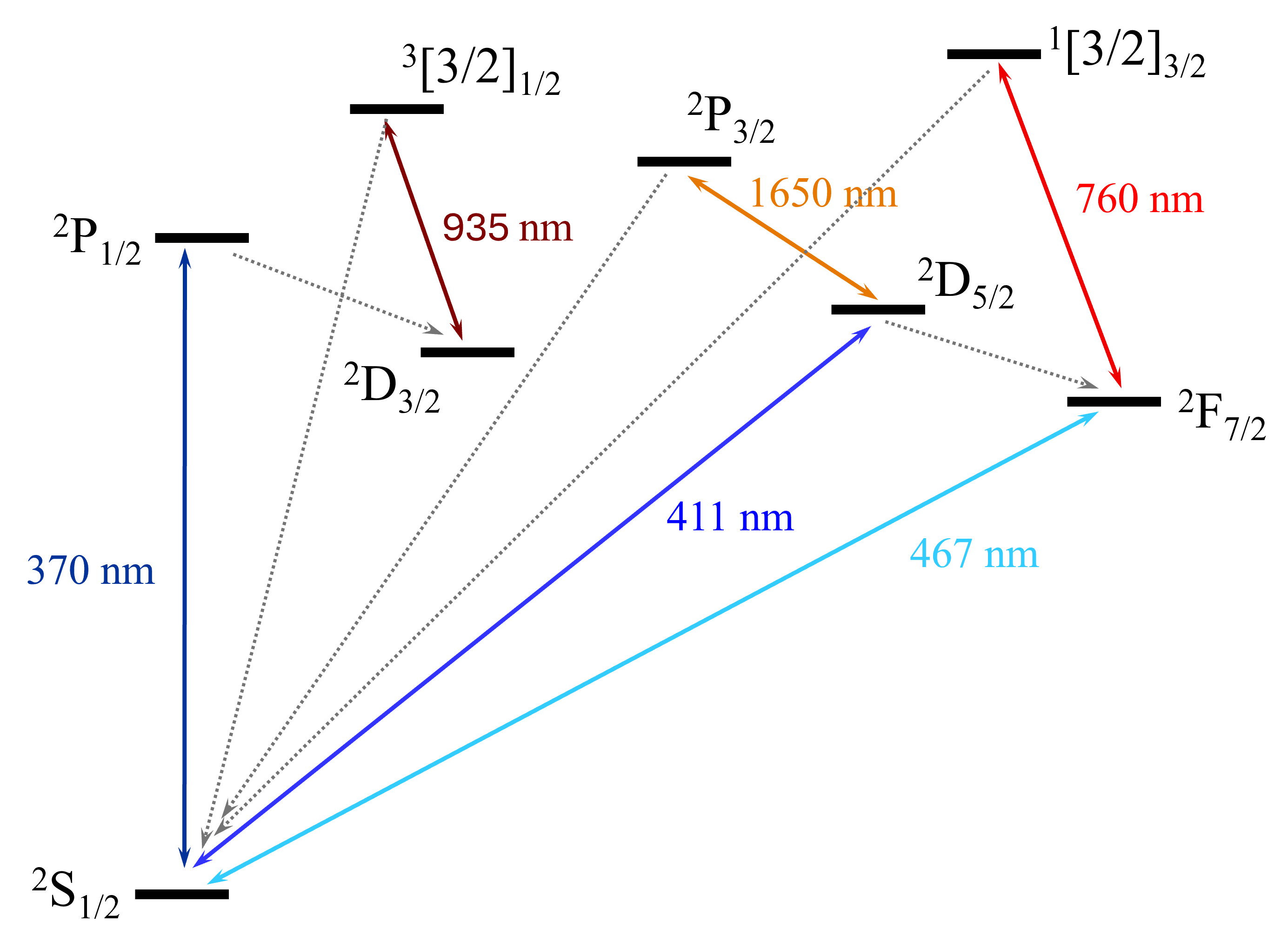}
	\caption{\label{fig:yb_scheme_simple}
		Simplified level scheme of $\mathrm{Yb^+}$, without hyperfine structure.}
        \label{fig:level_scheme_2}
\end{figure}

A general level scheme of \yb is shown in Fig.~\ref{fig:level_scheme_2}, which includes the $^2S_{1/2}\rightarrow ^2D_{5/2}$ electric quadrupole transition at 411 nm and the $^2D_{5/2}\rightarrow ^2P_{3/2}$ repumper transition at 1650 nm. The hyperfine mixing between the $^2P_{3/2}$ and the $^2F_{7/2}$ states is the dominant contribution to the HFE1 decay channel.

\section{E3 Laser setup}

\begin{figure}[h]
	\centering\includegraphics[width=0.4\textwidth]{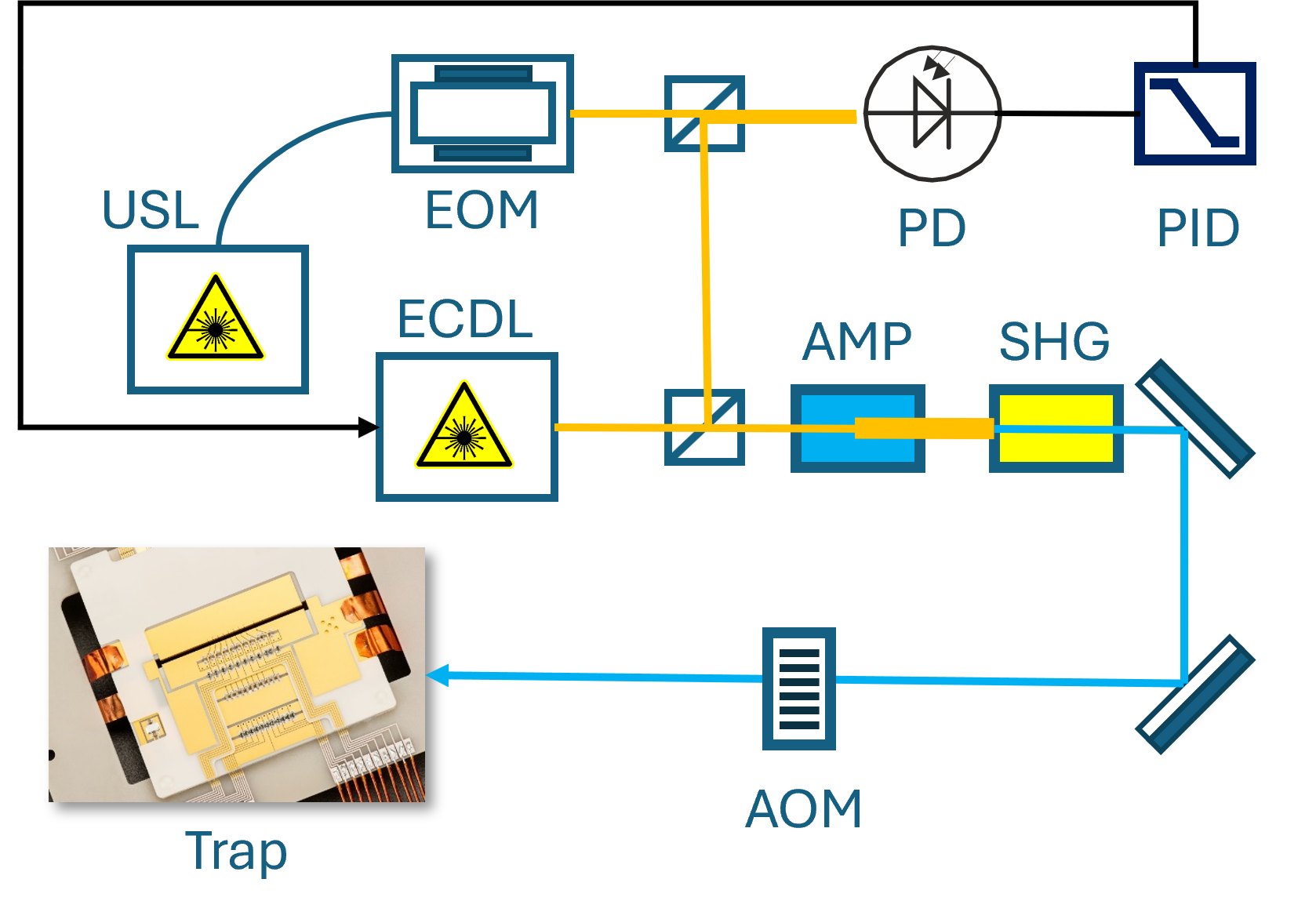}
	\caption{\label{fig:laser_setup}
		Scheme of the offset phase-locked loop of the spectroscopy laser (ECDL) to an ultra-stable reference laser (USL). Orange and blue lines indicate light at 934 nm and 467 nm respectively.}
\end{figure}

About 1 mW of light from an ultra-stable laser (USL) at 934 nm, used for the $^{171}\mathrm{Yb}^{+}$ ion clock~\cite{huntemannSingleIonAtomicClock2016a}, is delivered to our lab via an optical fiber.
This USL is stabilized to an optical resonator made of ultra-low expansion (ULE) glass and, for long-term stability, is referenced to a cryogenic silicon cavity~\cite{matei15LasersSub102017}, which drifts by 5 Hz/day at 1550 nm.
We use a fiber-EOM with a 5 GHz modulation bandwidth to generate sidebands at frequency $f_\mathrm{EOM}$ and offset phase-lock an external cavity diode laser (ECDL) to one of these sidebands of the USL (see Fig.~\ref{fig:laser_setup}).
The photodetector (PD) and the proportional-integral-differential (PID) controller are essential components for the offset phase-locked loop.
In this way, we bridge the isotope shift and the HF splitting between \ybe and \ybd. 
To ensure efficient second-harmonic generation (SHG), the offset phase-locked laser is further amplified using the injection-locking technique (AMP in Fig.~\ref{fig:laser_setup}).
The output of the injection-locked laser is sent through a periodically poled lithium niobate (PPLN) waveguide to generate the 467 nm light.
Fine-tuning of the 467 nm laser frequency is achieved using an acousto-optic modulator (AOM).
Thus, the frequency of the 467 nm laser can be tuned over a range of $\pm$~3 GHz with the EOM and $110\pm15$ MHz with the AOM.


\section{Measurement sequence for transition frequencies}\label{sec:meas_sequence}

\begin{figure}[h]
	\centering\includegraphics[width=0.48\textwidth]{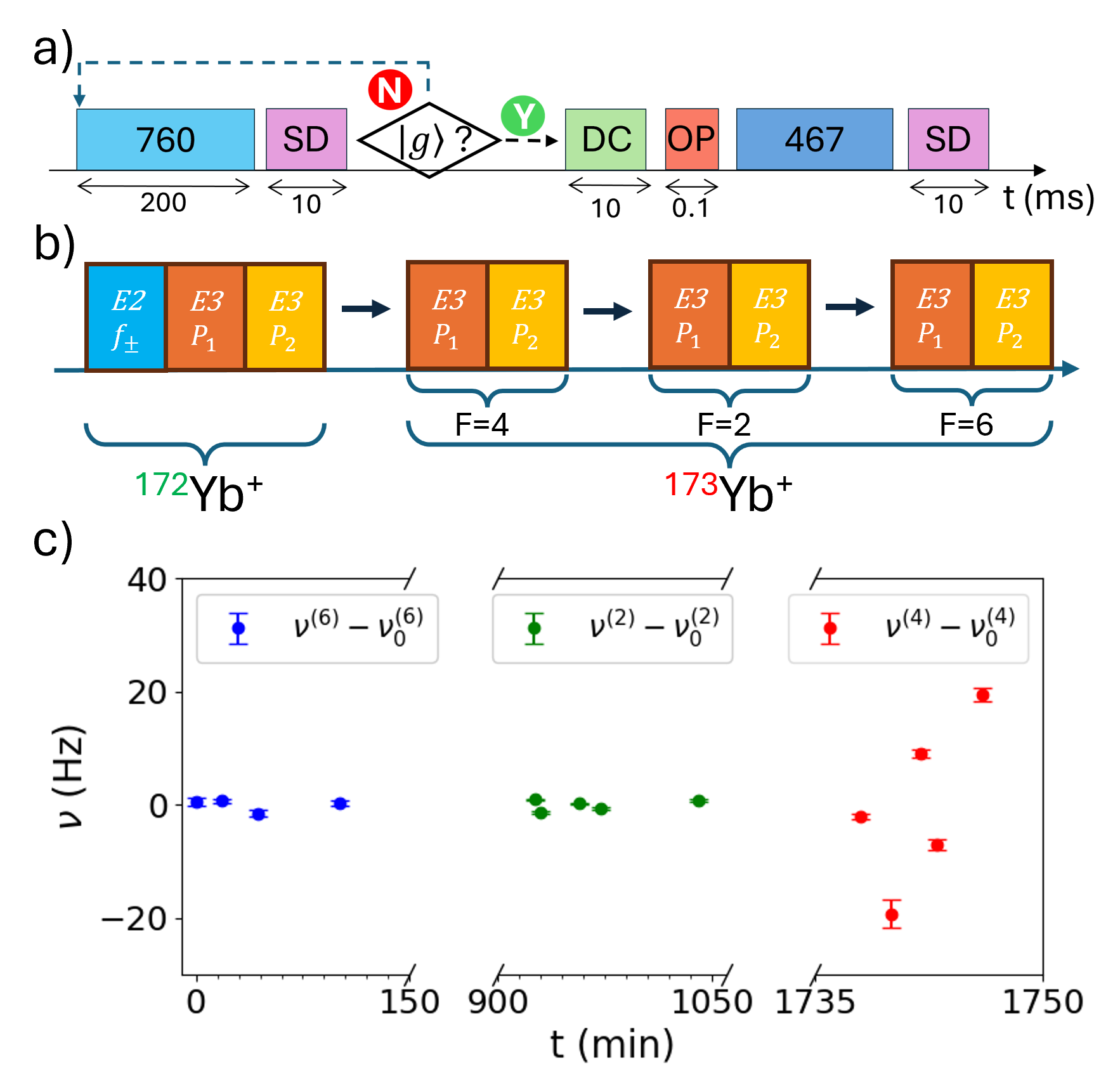}
	\caption{\label{fig:sequence_freq}
    The experimental sequence and the result of the absolute frequency measurements using the E3 transition of $^{172}$Yb$^{+}$ as a reference.
    a) The spectroscopy pulse sequence starts with 760 nm repumping followed by state detection (SD). These steps are repeated until the ion returns to the ground state $|g\rangle$. Then the ion is Doppler cooled (DC) and optically pumped (OP) to $|^2S_{1/2},F_g=3, m_e=0 \rangle$. Excitation by the 467 nm spectroscopy pulse is detected by SD. b) The experimental sequence used to determine the absolute transition frequencies. Blue box: spectroscopy on $^{172}\mathrm{Yb}^{+}$ E2 transitions for magnetic field measurement. Brown and orange boxes: spectroscopy pulse sequences used to record the spectra of the E3 transitions at optical power levels $P_1$ and $P_2$, respectively.
    c) The measured center frequencies at the power level $P_1$, subtracted from their average, for $F_e=6$ (blue), $F_e=2$ (green) and $F_e=4$ (red) with the fit uncertainty.
    }
\end{figure}

\begin{table*}[]
    \caption{\label{table:HF-splittings}
    The HF-splittings of $^2F_{7/2}$ state referenced to its $F_e=6$~HF-substate ($W_{F_e}-W_6$) from three measurement campaigns. The statistical ($u_A$) and systematic ($u_B$) uncertainties are shown in brackets, respectively. The dominant contributions to $u_A$ and $u_B$ are fitting errors and quadratic Zeeman shifts (QZS), respectively. To compare the transitions from different ground states, the ground state HF-splitting $\nu_\mathrm{hfs}=10491720239.55(9)$ Hz \cite{munchPreciseGroundstateHyperfine1987} and the QZS $\nu^{(2)}_\mathrm{QZS}=-\nu^{(3)}_\mathrm{QZS}=796(16)$ Hz are subtracted. We used all measurements to determine the frequencies of the five hyperfine splittings reported in the main text, together with their statistical and systematic uncertainties, evaluated using a conservative approach. The frequency values reported are obtained from the weighted average of the individual measurements. The statistical uncertainty is evaluated assuming uncorrelated measurements, while the systematic uncertainty is evaluated assuming fully correlated measurements.}
        \begin{ruledtabular}
            \begin{tabular}{clllc}
                Group & \textrm{$|^2S_{1/2}\rangle$}& \textrm{$|^2F_{7/2}\rangle$}& \textrm{$W_{F_e}-W_6$ (Hz)}  & RAP
                \\
                \colrule
               \#1 & $\mathrm{F_g=3,m_g=0}$ & $\mathrm{F_e=2,m_e=0}$ & \textrm{4 487 804 282 (2) (67)} & \textrm{N}\\
                  & $F_g=3,m_g=0$ & $F_e=4,m_e=0$ & 5 297 388 925 (7) (6820) & N\\
                \hline
               \#2 & $\mathrm{F_g=3,m_g=0}$ & $\mathrm{F_e=2,m_e=0}$ & \textrm{4 487 804 282 (3) (67)} & \textrm{N}\\
                  & $\mathrm{F_g=2,m_g=0}$ & $\mathrm{F_e=3,m_e=0}$ & \textrm{5 232 927 153 (4) (5989)} & \textrm{N}\\
                  & $F_g=3,m_g=0$ & $F_e=4,m_e=0$ & 5 297 389 030 (8)(6883) & N\\
                  & $\mathrm{F_g=2,m_g=0}$ & $\mathrm{F_e=5,m_e=0}$ & \textrm{3 884 874 663 (2) (198)} & \textrm{N}\\
            \hline
               \#3 & $\mathrm{F_g=2,m_g=0}$ & $\mathrm{F_e=1,m_e=0}$ & \textrm{3 659 258 444 (100) (577)} & \textrm{Y} \\                & $\mathrm{F_g=2,m_g=0}$ & $\mathrm{F_e=1,m_e=\pm1}$ & \textrm{3 659 258 495 (100) (429)} & \textrm{Y}\\
                & $\mathrm{F_g=3,m_g=0}$ & $\mathrm{F_e=3,m_e=\pm2}$ & \textrm{5 232 926 683 (100) (4590)} & \textrm{Y}\\
                & $F_g=3,m_g=0$ & $F_e=4,m_e=0$ & 5 297 389 104 (8) (6890) & N\\
                & $F_g=3,m_g=0$ & $F_e=4,m_e=\pm1$ & 5 297 389 174 (7) (6460) & N\\
                & $\mathrm{F_g=3,m_g=\pm3}$ & $\mathrm{F_e=4,m_e=\pm4}$ & \textrm{5 297 389 538 (500) (86)} & \textrm{Y}\\
                & $F_g=3,m_g=0$ & $F_e=5,m_e=\pm2$ & 3 884 874 583 (300) (152) & Y\\
            \end{tabular}
        \end{ruledtabular}
\end{table*}

The spectroscopy sequence of the \ybd clock transition is shown in Fig.~\ref{fig:sequence_freq}a.
As shown in Fig.~\ref{fig:sequence_freq}b, the experiment begins with a spectroscopy measurement on an $^{172}\mathrm{Yb}^{+}$ ion, where the beam pointing of the 467~nm E3 laser is optimized using the AC Stark shift it induces on the E2 transition at 411 nm, measured in a 5-ion $^{172}\mathrm{Yb}^{+}$ Coulomb crystal.
The ion number is then reduced to perform single-ion spectroscopy.
The $|^2S_{1/2},~m_J=\pm 1/2 \rangle \rightarrow |^2D_{5/2},~m_J=\pm 5/2 \rangle$ transitions are interrogated with 200 Hz accuracy to determine the magnetic field (light blue box).
After switching the fiber electro-optic modulator (EOM) frequency, the $|^2S_{1/2},~m_J=-1/2 \rangle \rightarrow |^2F_{7/2},~m_J=-1/2 \rangle$ E3 transition is measured at two different clock laser intensities (brown and orange boxes) to extract the AC Stark shift.
Previously, we measured the ratio of the first-order Zeeman sensitivities of the $^{172}\mathrm{Yb}^{+}$ E3 and E2 transitions to be $\delta\nu^{(\mathrm{E3})}/\delta\nu^{(\mathrm{E2})}=0.21547(2)$.
Using this ratio, we calculate the center frequency of the $^{172}\mathrm{Yb}^{+}$ E3 transition, extrapolated to zero AC Stark shift. The laser frequencies are then switched to \ybd. The E3 laser frequency is shifted via the fiber EOM and double-pass acousto-optic modulator (AOM), and the transition frequencies relative to $\nu_{172}$ are determined from the known modulator frequencies.

\begin{table*}
\caption{\label{table:zeeman} First-order $\mu_1$ and second-order $\mu^{(m)}_2$ Zeeman sensitivities for the $^{173}\mathrm{Yb}^{+}$ $^2S_{1/2}$ and $^2F_{7/2}$ states, calculated from experimentally determined hyperfine splittings.}
\begin{ruledtabular}
\begin{tabular}{llrrrrrrrrr}
\multicolumn{2}{c}{\textrm{State}} & \textrm{$\mu_1$ (MHz/mT)}& \multicolumn{7}{c}{\textrm{$\mu^{(m)}_2~\mathrm{(mHz/\mu T^2)}$}}\\
 &  &  & $m_e=0$ & $m_e=1$ & $m_e=2$  & $m_e=3$  & $m_e=4$
 & $m_e=5$ & $m_e=6$
\\
\hline
$^2S_{1/2}$ & $F_e=2$ & -4.7 &  18.7 & 16.6 & 10.4 & & & &  &
 \\
&$F_e=3$ & 4.7 & -18.7 & -16.6 & -10.4 & 0 & & &  &\\
\hline
$^2F_{7/2}$ & $F_e=1$ & 36.0  & -694.8 & -521.1 & & & &  &
 \\
&$F_e=2$ & 17.3 &  -90.1 & -176.6 & -436.0 & & &  &
 \\
&$F_e=3$ & 12.7 &  -7011.8 & -6611.7 & -5411.5 & -3411.0 &  &  & 
 \\
&$F_e=4$ & 10.8 &  8060.2 & 7562.3 & 6068.8 & 3579.7 & 94.9 &  & 
 \\
&$F_e=5$ & 9.9 &  -211.1 & -202.0 & -174.8 & -129.3 & -65.7 & 16.0 & 
 \\
& $F_e=6$ & 9.3 & -52.4 & -50.9 & -46.6 & -39.3 & -29.1 & -16.0 & 0
 \\
\end{tabular}
\end{ruledtabular}
\end{table*}

\begin{figure}[h]
	\centering\includegraphics[width=0.48\textwidth]{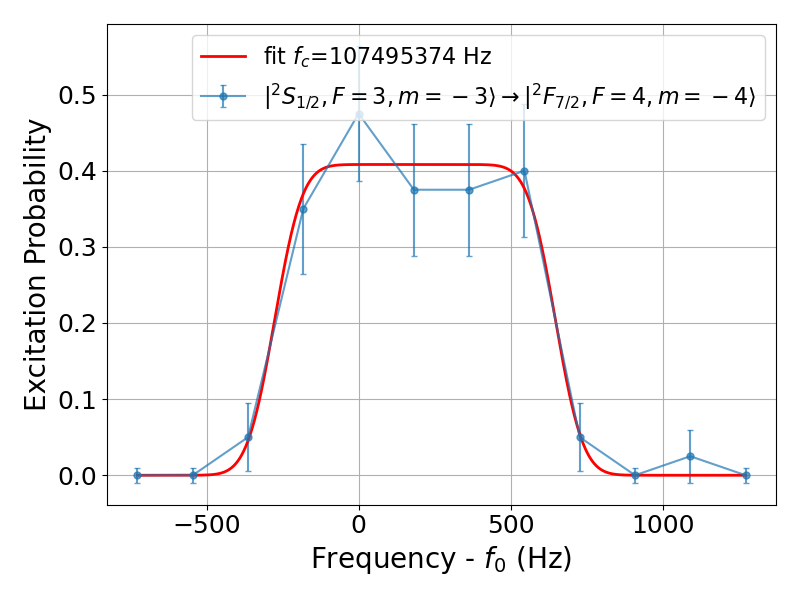}
	\caption{\label{fig:spectra_RAP}
    Experimentally recorded spectra of the $|^2S_{1/2},F_g=3,m_g=-3\rangle$ to $|^2F_{7/2},F_e=4,m_e=-4\rangle$ transition, interrogated by the RAP technique. The experimental data is fitted with the convolution of Gaussian and rectangular functions.
    }
\end{figure}

The transition frequencies to the $F_e = 2, 4$ and 6 hyperfine states are measured at two different optical powers to extrapolate the AC Stark shift.
For each $F_e$ state, five spectra are recorded at the higher power level $P_1$, with a typical resolution at the Hz level. At the lower power level $P_2$ ($P_2/P_1\approx0.25$ with $<0.002$ uncertainty), only one or two spectra are taken, as the narrower linewidth already permits determination of the center frequency with Hz-level precision.
 Figure~\ref{fig:sequence_freq}c shows the frequency stability of the recorded spectra. The $F_e=4$ transition exhibits significantly more scatter due to its strong quadratic Zeeman shift (QZS), in contrast to the Zeeman-insensitive $F_e=2$ and $F_e=6$ transitions.
From this dataset, we determine the absolute frequencies of the $|^2S_{1/2}, F_g = 3\rangle \rightarrow |^2F_{7/2}, F_e = 2, 4, 6\rangle$ transitions to be:
$\nu^{(2)}~=~642.123664367825(16)(434)$~THz,
$\nu^{(4)}~=~642.124473952467(17)(6834)$~THz, and
$\nu^{(6)}~=~642.119176563542(16)(431)$~THz,
where the numbers in parentheses denote the statistical and systematic uncertainties, respectively.
The measured hyperfine splittings are summarized in Table~\ref{table:HF-splittings}, denoted as group~\#1.
Due to the large quadratic Zeeman sensitivity (see Table~\ref{table:zeeman}), the uncertainty on $\nu^{(4)}$ is the largest.

To determine additional hyperfine splittings, we installed a radio-frequency (RF) antenna, enabling state preparation in the $|^2S_{1/2}, F_g = 2, m_g = 0\rangle$ state.
The results are summarized in Table~\ref{table:HF-splittings}, marked as group \#2.  

To reduce the systematic uncertainty arising from the quadratic Zeeman shift, several first-order Zeeman-sensitive transitions are interrogated using the RAP technique.
The recorded spectra are fitted with the convolution of Gaussian and rectangular functions. 
A typical spectrum and its fit are shown in Fig. \ref{fig:spectra_RAP}.
By taking the average of the transitions to opposite Zeeman substates, the first-order Zeeman shift is removed.
The averaged frequency is then corrected for the QZS using the sensitivities listed in Table \ref{table:zeeman} and the magnetic field measured via the \ybz E2 transition.

Transition frequencies of \ybd are referenced to the coherently driven $|^2S_{1/2}, F_g=3\rangle \rightarrow |^2F_{7/2}, F_e=6\rangle$ transition, assuming the RAP-interrogated transitions have the same Stark shift sensitivity as this reference transition.
The measured hyperfine splittings are listed in Table~\ref{table:HF-splittings}.

\section{Zeeman sensitivity of \ybd~$^2F_{7/2}$ hyperfine states}\label{sec:zeeman_sensitivity}
The total Zeeman shift can be calculated as $\Delta\nu_\mathrm{Zm}~=~\mu_1 m B+\mu_2^{(m)} B^2$, where $\mu_1=\mu_B g_F/h$ is the first-order Zeeman sensitivity and the second-order sensitivity, defined as:
\begin{equation}
   \mu_2^{(m)}=\frac{\mu_B^2(g_J-g_I)^2}{h}\sum_{\Delta F=\pm1}\frac{| \langle \gamma F m| J_z| \gamma F' m \rangle  |^2}{\Delta \nu_{F' F}} .
\end{equation}
 The magnetic field sensitivities of all states of the $^{173}\mathrm{Yb}^{+}$ $^2F_{7/2}$-manifold are summarized in Table \ref{table:zeeman}.

\section{Leading systematics of the transition frequency}

The leading contributions to the measured absolute E3 transition frequencies are summarized in Table \ref{table:freq_shift}.
The uncertainty from the reference transition, i.e., the \ybz~E3 transition, is dominated by magnetic field fluctuations of 0.07 $\mu$T in the lab. To calculate the quadratic Zeeman shift (QZS), both the quadratic Zeeman sensitivity and the absolute magnetic field strength are required.
\begin{table}[h]
    \caption{
    Leading frequency shifts $\delta\nu$ and corresponding uncertainties $\sigma_\nu$ in Hz for the absolute E3 transition frequencies (Table I in main text) from the $|^2S_{1/2}, F_g=3, m=0\rangle$ state to the $|^2F_{7/2}, F_e=2,6, m=0\rangle$ states in \ybd.\label{table:freq_shift}
    }
\begin{ruledtabular}
\begin{tabular}{lrrrr}
& \multicolumn{2}{c}{\textrm{$F_e=2$}}&  \multicolumn{2}{c}{\textrm{$F_e=6$}}\\
\textrm{Effect} &  $\delta\nu$ & $\sigma_\nu$ &  $\delta\nu$ & $\sigma_\nu$ 
\\
\hline
 \textrm{Ref. transition} & 2486276 &  432 &  2486276 & 432
\\
 \textrm{AC Stark shift} & 172.58 &  0.29  & 168.55 & 0.14
 \\
 \textrm{QZS} & -3021 & 60  & -1426 & 29 
 \\
 \textrm{BBR}\footnote{assuming a similar differential polarizability as \ybe within 100\% uncertainty} & -0.046 & 0.098  & -0.046 & 0.098
 \\
  \textrm{Quadrupole shift} & -0.019 & 0.003  & -0.019 & 0.003
\\
 \hline
 \textrm{Total} & 2483427.52 & 435 & 2485018.49 & 432 \\
\end{tabular}
\end{ruledtabular}
\end{table}
The quadratic Zeeman sensitivities of the \ybd~$^2F_{7/2}$ HF states are calculated from the measured HF splittings (see SM \cite{yuSupplementaryMaterialThis2024}).
Following the uncertainty analysis in \ybe measurements \cite{hosakaOpticalFrequencyStandard2005, huntemannHighaccuracyOpticalClock2014a}, we estimate a 1\% uncertainty in the calculated sensitivities.
The absolute magnetic field strength is calculated from the measured Zeeman splitting of the \ybz electric quadrupole transitions with an accuracy of $5\times10^{-3}$, limited by the uncertainty in the $g$-factor \cite{meggersSecondSpectrumYtterbium1967, gosselCalculationStronglyForbidden2013}.

\section{Measurement sequence for Rabi frequency}

\begin{figure}[h]
	\centering\includegraphics[width=0.48\textwidth]{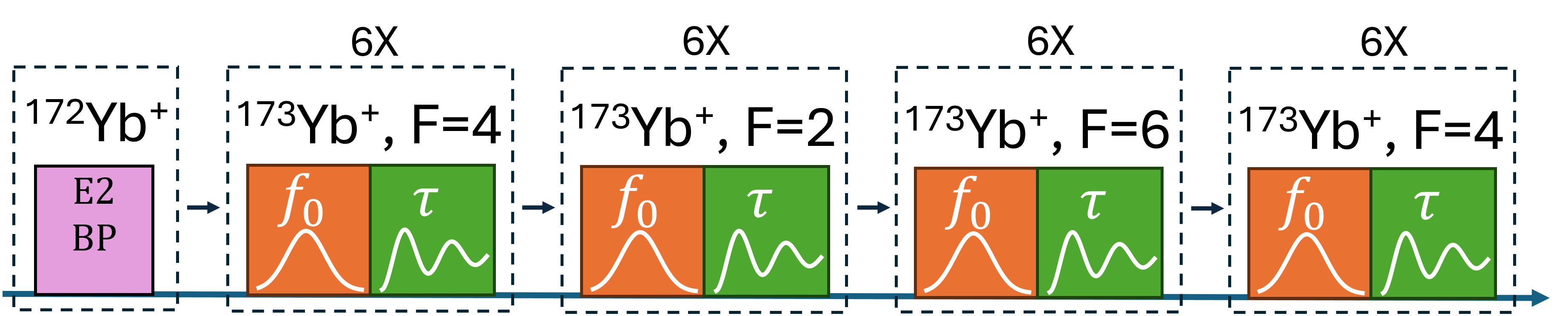}
	\caption{\label{fig:sequence1}
	Experimental sequence to determine the Rabi frequency.
Purple box: spectroscopy on $^{172}\mathrm{Yb}^+$ to measure the magnetic field and align the E3 laser beam pointing (BP).
Orange box: record spectra of the transition.
Green box: record Rabi oscillation of the transition.
}
\end{figure}

The measurement sequence for the Rabi frequency begins with spectroscopy on $^{172}\mathrm{Yb}^+$ (purple box, Fig.~\ref{fig:sequence1}) to align the E3 laser beam. Then, the laser frequencies are switched to \ybd, and the three hyperfine states are measured in the order shown in Fig.~\ref{fig:sequence1}.
In the experiment, a nonzero laser detuning $\Delta$ can shift the measured effective Rabi frequency to $\Omega_{\mathrm{eff}} = \sqrt{\Omega_{0}^2 + \Delta^2}$.
To ensure $\Delta \approx 0$ (and thus $\Omega_\mathrm{eff} = \Omega_{0}$), each dataset includes 6 Rabi oscillation measurements (green box) interleaved with frequency scans (orange box), minimizing possible frequency detunings due to laser frequency drift and magnetic field fluctuations. During the measurements, the E3 laser power is actively stabilized and monitored using an out-of-loop photodetector.

\section{Discussion of the $\Omega^{(F_g \rightarrow F_e)}$ fit model}
As described in the main text, $\Omega^{(F_g \rightarrow F_e)}$ is derived by fitting coherent Rabi flops at identical laser power.
The uncertainty arising from the E3 laser beam pointing is accounted for by repeating the measurements and performing statistical analysis to the results.
Decoherence is another effect that could effectively change $\Omega^{(F_g \rightarrow F_e)}$.
Possible contributions to decoherence are thermal decoherence, magnetic field fluctuations and laser phase noise.  

Thermal decoherence is described using the Jaynes-Cummings (JC) model, where the total Rabi oscillation is governed by different thermally excited states with individual Rabi frequencies, such as $p(t) = p_0\sum\limits_{n=0}^{\infty} \frac{\bar{n}^n}{(1+\bar{n})^{n+1}}\sin(\Omega^{'}_n t/2)^2$, with $p_0$ the maximum excitation probability. Here the Rabi frequency of a trapped ion $\Omega^{'}_n$ is related to the Rabi frequency of a free atom $\Omega_0$ through the expression $\Omega^{'}_n=\Omega_0 L_n(\eta^2)e^{-\eta^2/2}$, where $\eta$ is the Lamb-Dicke parameter, and $L_n(x)$ is the Laguerre polynomial.
In our experiment, we rotate the radial principal axes by fine-tuning the trap compensation voltage so that only one secular motion ($\omega_{x}=2\pi\times650$ kHz) has a projection on the E3 spectroscopy laser.
The Doppler temperature of 0.5 mK at the dipole allowed cooling transition of $\mathrm{Yb}^+$ equals to a mean phonon number of $\bar{n}=15$. The Lamb-Dicke parameter of the ground state is $\eta=0.09$. 

We model magnetic field fluctuations and laser phase noise with an exponential decay (EXP model), such as $p(t)=p_0 \times \sin{\Omega_{0} t}\times e^{-t/t_\mathrm{decoh}}$, where $t_\mathrm{decoh}$ is the decoherence time.

We can combine both models (JC + EXP) to define the excitation probability as $p(t)=p_0\left(\sum\limits_{n=0}^{\infty} \frac{\bar{n}^n}{(1+\bar{n})^{n+1}}\sin(\Omega^{'}_n t/2)^2\right)e^{-t/\tau_{\mathrm{decoh}}}$.

\subsection{Measurement campaign at low $\bar{n}$}

In our experiment, factors that could lead to ion temperature changes are 1) 370 nm laser power fluctuations and 2) the temperature-sensitive modulation depth of the EOM.
To maintain a stable ion temperature during the measurement described in the main text, where decoherence times were determined, the experiment was conducted on a Sunday to minimize laboratory temperature fluctuations. 
Additionally, throughout the total measurement duration of 140 minutes, the optical power of both 370 nm beams was carefully monitored and fine-tuned every 30 minutes to suppress power fluctuations to below 1\%.
In this measurement, if the observed decoherence arises solely from thermal dephasing (which is not strictly accurate), the decoherence would correspond to ion temperatures of $\bar{n} = 17.6(1.4)$, $26.7(2.8)$ and $17.6(1.4)$ for the $|^2S_{1/2}, F_g=3\rangle~\rightarrow~|^2F_{7/2},F_e~=~2,4,6 \rangle$ transitions respectively.
The significantly higher $\bar{n}$ for the transition to the $F_e=4$ state confirms the presence of an additional decoherence source.
In contrast, the lower fitted temperatures for the $F_e=2$ and $F_e=6$ transitions are close to the Doppler limit ($\bar{n} = 15$), supporting the assumption that all ions were cooled to the Doppler temperature. Under this assumption, we extract the corresponding decoherence times to be
$\tau_{\mathrm{decoh}}~=~1777(865)$~ms and $108(22)$~ms, and $2370(1425)$ ms for transitions to $F_e = 2, 4, 6$, respectively.
Given that the transition to $F_e=4$ has a quadratic Zeeman sensitivity two orders of magnitude stronger than the other transitions, and given that all transitions shared the same cooling and initial state preparation, we attribute the shorter decoherence time $\tau_{\mathrm{decoh}}~=~108(22)$~ms for $F_e=4$ to magnetic field fluctuations.
From this measurement, the Rabi frequency ratios are determined to be $\Omega^{(3\rightarrow2)}/\Omega^{(3\rightarrow6)}= 0.997(35)$ and $\Omega^{(3\rightarrow4)}/\Omega^{(3\rightarrow6)} = 4.34(16)$, where the uncertainty includes the fitting error and the laser intensity fluctuations.

\subsection{Measurement campaign at increased $\bar{n}$}

\begin{figure}[h]
	\centering\includegraphics[width=0.45\textwidth]{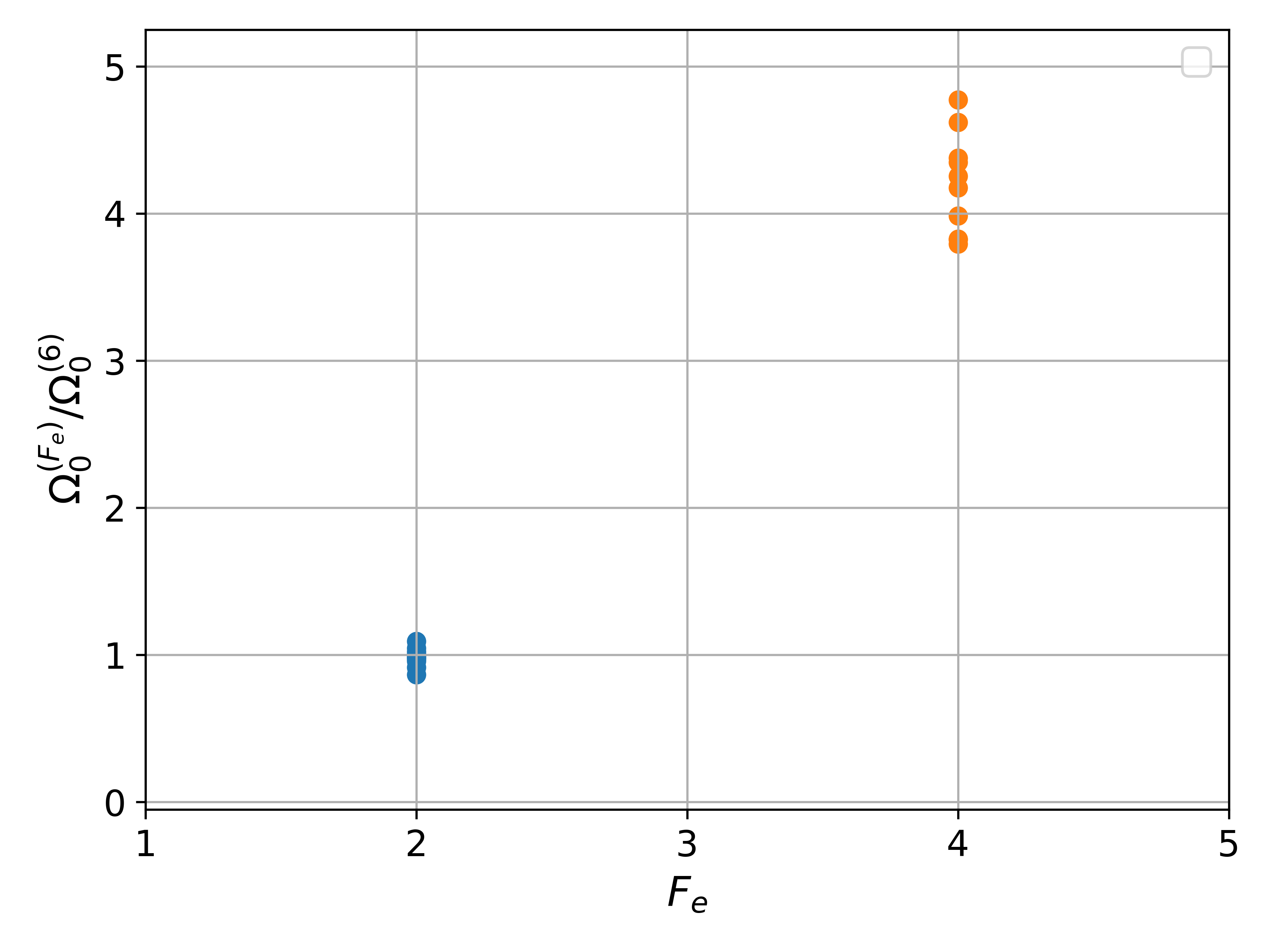}
	\caption{\label{fig:omega_ratio}
	Rabi frequency ratio between the quenched transition and the reference transition, determined from thermal dephasing model (JC). The error bars from the fit are not visible at this scale. }
\end{figure}

\begin{figure}
	\centering\includegraphics[width=0.45\textwidth]{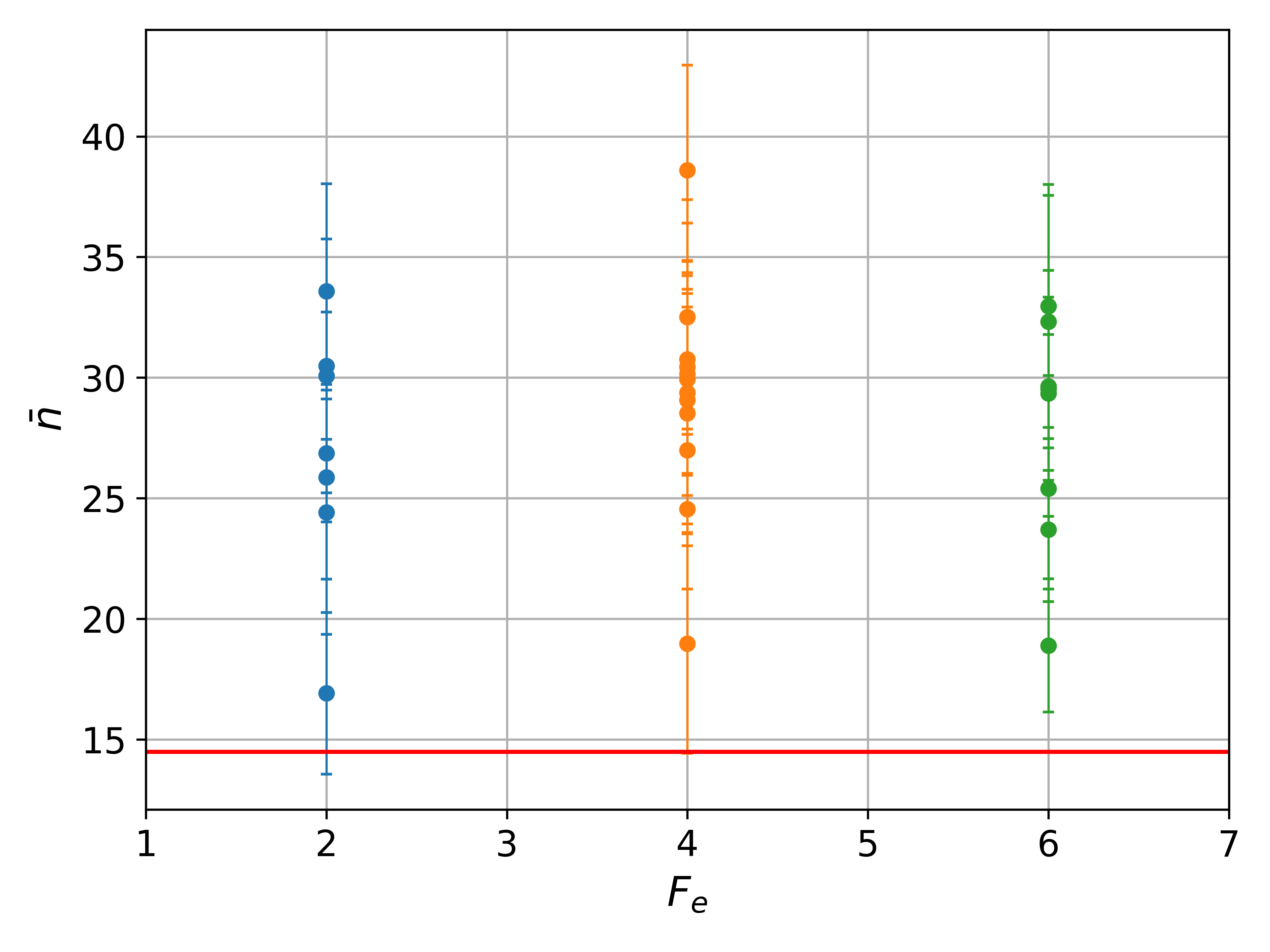}
	\caption{\label{fig:nbar}
	Mean phonon number $\bar{n}$ and the fit error for the transitions to different $^2F_{7/2}$ hyperfine states in the second measurement campaign, extracted from experimental Rabi oscillations fitted with JC+EXP1 model. The coherence time of the individual transitions are assumed to be the same as in the first measurement campaign. The Doppler limit is indicated by a red horizontal line.}
\end{figure}

To validate our model, we performed a second measurement campaign with increased ion temperature, achieved by a heating beam at 370 nm. A total of 30 datasets were collected.
Assuming a constant decoherence time identical to that of the previous measurement, we used the JC model to extract the Rabi frequency $\Omega$ and the mean motional quantum number $\bar{n}$.
The resulting values of $\Omega^{(F_g\rightarrow F_e)}/\Omega^{(3\rightarrow6)}$ and $\bar{n}$ are shown in Fig.~\ref{fig:omega_ratio} and Fig.~\ref{fig:nbar}, respectively.

To evaluate the robustness of our results, we consider four scenarios, in which the decoherence 1) is solely caused by thermal dephasing (JC), 2) is solely caused by magnetic field fluctuations (EXP), 3) arises from both thermal dephasing and magnetic field fluctuations, with fixed decoherence times taken from the pre-measurement (JC+EXP1) and 4) similar to 3) without specifying the decoherence time (JC+EXP2).
The results are summarized in Table~\ref{table:compare_model}.
The choice of the fitting model and fit parameters leads to only marginal differences within the uncertainties, confirming the validity of our approach. We take the values from JC+EXP1 model for our final result.

Independent of the fit model, the $\Omega^{(F_g\rightarrow F_e)}/\Omega^{(3\rightarrow6)}$ ratios for the two E3 transitions exhibit comparable relative uncertainties of about 7\%.
While the uncertainty from curve fitting is only 2\%, the dominant contribution to the uncertainty in this measurement originates from the E3 beam pointing.
This is confirmed by analyzing the variation in $\Omega^{(F_e)}$ for the same transition at different times, which results in an uncertainty of 7\%.
Such variation corresponds to a 12\% change in beam intensity, attributed to the small beam waist ($30~\mathrm{\mu m}$) and the large distance (24 cm) between the final mirror and the ion.

In the main text, we report the averaged $\Omega^{(F_g\rightarrow F_e)}/\Omega^{(3\rightarrow6)}$ ratios, combining uncertainties from both measurement campaigns.

\begin{table}[]
\caption{\label{table:compare_model}Experimentally determined $\Omega^{(F_g\rightarrow F_e)}/\Omega^{(3\rightarrow6)}$ using Jaynes-Cumming (JC), a sine function with exponential decay (EXP) and the combined model with and without specifying the $t_\mathrm{decoh}$ (JC+EXP1 and JC+EXP2). Numbers in the brackets represents the statistical uncertainty.}
\begin{ruledtabular}
\begin{tabular}{lcccc}
HF-State & JC &
EXP & JC+EXP1 & JC+EXP2 \\
\hline
$F_e$=2 &  0.98(6) &  1.00(4)  & 0.98(7) &  0.99(6) \\
$F_e$=4 &   4.16(30) &  4.03(28)   & 4.09(37) & 4.08(31) \\
\end{tabular}
\end{ruledtabular}
\end{table}

\section{Hyperfine-induced E1 Transition}

In \ybd, the extremely weak E3 clock transition $^1S_{1/2} \rightarrow  ^2F_{7/2}$ is enhanced by the large electric quadrupole moment of its deformed nucleus, giving rise to a hyperfine-induced electric dipole (HFE1) transition that involves mixing with intermediate states \cite{dzubaHyperfineInducedElectricDipole2016}. 

\begin{figure}[ht]
\begin{center}
  \includegraphics[width=0.45\textwidth]{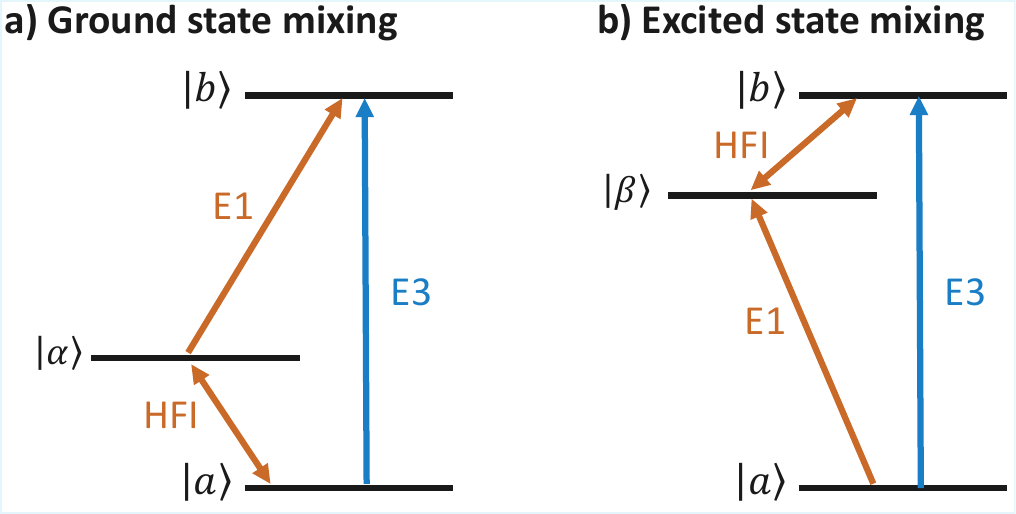}
   \caption{Simplification of the hyperfine-induced E1 (HFE1) mechanism, illustrated by the orange paths. It shows how HFE1 coupling opens a channel that interferes with the weak E3 interaction between $|a\rangle$ and $|b\rangle$ through two contributions: \textbf{a)} State $|a\rangle$ mixes with an intermediate state $|\alpha\rangle$ via hyperfine interaction (HFI). Since $|\alpha\rangle$ has E1 coupling to $|b\rangle$, the transition $|a\rangle \leftrightarrow |b\rangle$ acquires an E1 component.
    \textbf{b)} Similarly, state $|b\rangle$ mixes with an intermediate state $|\beta\rangle$, which couples via E1 to $|a\rangle$.}
  \label{fig:HFI_E1}
\end{center}
\end{figure}

According to Eq.~1 in Ref.\cite{dzubaHyperfineInducedElectricDipole2016}, the amplitude of the HFE1 transition is given by:

\begin{eqnarray}
 A_{\mathrm{HFE1}}(b \rightarrow a) &=&  \sum_n  \Biggl[\frac{\langle a|\hat{H}_\mathrm{HFI}|n\rangle \langle n|\hat{D}|b \rangle }{E_a - E_n} \nonumber\\
 &&+ \frac{\langle b|\hat{H}_\mathrm{HFI}|n\rangle \langle n|\hat{D}|a\rangle}{E_b - E_n} \Biggl], 
 \label{eq_hfssum}
\end{eqnarray}
 where $\hat{H}_\mathrm{HFI}$ is the hyperfine interaction Hamiltonian, $\hat{D}$ is the dipole operator, and $\ket*{n}$ is an intermediate state that can couple to either $\ket*{a}$ or $\ket*{b}$ via an electric dipole (E1) transition. Eq.~\ref{eq_hfssum} can be interpreted intuitively, the two terms represent distinct physical contributions in which the sums over intermediate states are independent, as seen in:
\begin{eqnarray}
 A_{\mathrm{HFE1}}(b \rightarrow a) &=& \sum_\alpha\frac{\langle a|\hat{H}_\mathrm{HFI}|\alpha\rangle \langle \alpha|\hat{D}|b\rangle}{E_a - E_\alpha} \nonumber \\
 && +\sum_\beta \frac{\langle b|\hat{H}_\mathrm{HFI}|\beta\rangle \langle \beta|\hat{D}|a\rangle }{E_b - E_\beta}.
 \label{eq_hfssum2}
\end{eqnarray}
Figure \ref{fig:HFI_E1} illustrates the two distinct contributions:
\begin{itemize}
\item \textbf{Ground state mixing:} the hyperfine interaction, $\matrixel*{a}{\hat{H}_\mathrm{HFI}}{\alpha}$, causes the intermediate state $\ket*{\alpha}$ to \textbf{mix} with $\ket*{a}$. Since the state $\ket*{\alpha}$ also couples to $\ket*{b}$ via an E1 transition, the transition $\ket*{a} \rightarrow \ket*{b}$ has an E1 contribution.
\item \textbf{Excited state mixing:} the hyperfine interaction, $\matrixel*{b}{\hat{H}_\mathrm{HFI}}{\beta}$, causes the intermediate state $\ket*{\beta}$ to \textbf{mix} with $\ket*{b}$. Since the state $\ket*{\beta}$ also couples to $\ket*{a}$ via an E1 transition, the transition $\ket*{a} \rightarrow \ket*{b}$ has an E1 contribution.
\end{itemize}

In the case of the $^{173}\mathrm{Yb}^+$ E3 transition, we express the clock and ground states perturbed by the hyperfine interaction, using second-order perturbation theory:

\begin{eqnarray}
\label{eq_statemix}
\ket{\tilde{S}_{1/2}} &=& \ket{S_{1/2}} + \sum_\alpha\frac{\langle \alpha|\hat{H}_\mathrm{HFI}|S_{1/2}\rangle}{E_{S_{1/2}}-E_\alpha}|\alpha\rangle \nonumber\\
&\\
\ket{\tilde{F}_{7/2}}  &=& \ket{F_{7/2}} + \sum_\beta\frac{\langle\beta|\hat{H}_\mathrm{HFI}|F_{7/2}\rangle}{E_{F_{7/2}}-E_\beta}|\beta\rangle \nonumber,
\end{eqnarray}
where  the first terms denote the unperturbed states without hyperfine mixing, and the second terms denote the correction induced by the hyperfine interaction. The summations formally include all states that can couple via the hyperfine interaction, but in practice, only those with allowed E1 coupling to either $\ket{S_{1/2}}$ or $\ket{F_{7/2}}$ contribute to the HFE1 effect, as seen in Fig. \ref{fig:HFI_E1}.

To identify which intermediate states actually contribute, we now express the matrix element of the hyperfine interaction $\hat{H}_\mathrm{HFI}$:

\begin{eqnarray}
&\langle\gamma' I J' F' m'_F|\hat{H}_\mathrm{HFI}|\gamma I J F m_F\rangle =\delta_{F' F}\delta_{m'_F m_F}(-1)^{I+J+F} \nonumber\\
&\times \sum_k
\left\{\begin{matrix}
F & J & I\\
k & I & J'
\end{matrix}\right\}\langle \gamma' J'||T^e_k||\gamma J\rangle \langle I||T^n_k||I\rangle,
\label{eq_HFSmxel}
\end{eqnarray}
where $\gamma$, $\gamma'$ are other quantum numbers that label the states and $k$ is the tensor rank determined by $2^k$-moment of the nucleus. 
The Kronecker deltas $\delta_{F' F}$ and $\delta_{m'_F m_F}$ in Eq.~\ref{eq_HFSmxel} imply that hyperfine mixing can only occur between states with the same total angular momentum $F$ and magnetic quantum number $m_F$.
For example, all intermediate states mixing with $\ket*{S_{1/2},F_g=3,m_g=0}$ must also have $(F=3, m_F=0)$.
Combining these hyperfine mixing selection rules and the selections rules of an E1 transition, it follows that the states $\ket*{F_{7/2},F_e=5,6}$ cannot acquire HFE1 coupling.

\section{Rabi frequency and transition rates}
   
The amplitude of the HFE1 transition can be linked to its transition rate as follows \cite{dzubaHyperfineInducedElectricDipole2016}:
\begin{equation}
\mathcal{R}_{\mathrm{HFE1}}^{(F_e\rightarrow F_g)} = \frac{4}{3} \frac{c \alpha k^3}{e^2}
\frac{A_{\mathrm{HFE1}}^2}{2F_e+1},
\end{equation}
where $A_{\mathrm{HFE1}} = \langle F_e \,\|\, e\hat{r} \,\|\, F_g\rangle$ is defined in Eq. \ref{eq_hfssum}.

The corresponding Rabi frequency for this HFE1 transition (excited by $\pi$-polarized light) is given by:  
\begin{equation}
\Omega_{\mathrm{HFE1}}^{(F_g\rightarrow F_e)} = \pm K_{\mathrm{E1}}^{(F_e\rightarrow F_g)} \sqrt{\mathcal{R}_{\mathrm{HFE1}}^{(F_e\rightarrow F_g)}},
\label{omega_HFE1}
\end{equation}
where the $\pm$ sign accounts for the phase of the transition amplitude and the coefficient $K_{\mathrm{E1}}^{(F_e\rightarrow F_g)}$ is defined as:
\begin{equation}
K_{\mathrm{E1}}^{(F_e\rightarrow F_g)} = \frac{eE_0}{\hbar} \sqrt{\frac{3}{4c\alpha k^3}}~\langle F_g m_g, 1,0 \mid F_e m_e\rangle,
\label{K_E1}
\end{equation}
and $\langle F_g m_g, 1,0 \mid F_e m_e \rangle$ is the Clebsch-Gordan coefficient. 

Similarly, for a pure E3 transition excited by $\pi$-polarized light is
\begin{equation}
\Omega_{\mathrm{E3}}^{(F_g\rightarrow F_e)} = K_{\mathrm{E3}}^{(F_e\rightarrow F_g)} \sqrt{\mathcal{R}_{\mathrm{E3}}^{(F_e\rightarrow F_g)}},
\end{equation}
with

\begin{equation}
K_{\mathrm{E3}}^{(F_e\rightarrow F_g)} = \frac{eE_0}{\hbar}~ \sqrt{\frac{21}{32 c\alpha k^3}}
~\langle F_g m_g, 3,0 \mid F_e m_e \rangle.
\end{equation}

For quenched transitions such as $|^2S_{1/2}, F_g = 3\rangle \rightarrow |^2F_{7/2}, F_e = 2,4\rangle$, the total observed Rabi frequency combines both contributions:
\begin{equation}
\Omega^{(F_g\rightarrow F_e)} = \abs{\Omega^{(F_g\rightarrow F_e)}_{\mathrm{HFE1}}+\Omega_{\mathrm{E3}}^{(F_g\rightarrow F_e)}}
\label{eq_sum_omega}
\end{equation}
Based on the experimental result that $\Omega^{(F_g\rightarrow F_e)}>\Omega^{(F_g\rightarrow F_e)}_{\mathrm{E3}}$, and substituting the expressions above:
\begin{equation}
\Omega^{(F_g\rightarrow F_e)} = K_{\mathrm{E1}}^{(F_e\rightarrow F_g)} \sqrt{\mathcal{R}_{\mathrm{HFE1}}^{(F_e\rightarrow F_g)}}
\pm K_{\mathrm{E3}}^{(F_e\rightarrow F_g)} \sqrt{\mathcal{R}_{\mathrm{E3}}^{(F_e\rightarrow F_g)}}
\end{equation}

Here, the $\pm$ arises from the fact that $A_{\mathrm{HFE1}}$ can have either sign, unlike a pure E1 (or pure E3) matrix element where the sign does not affect the measured Rabi frequency. This sign ambiguity comes from the coherent sum over hyperfine mixing contributions (see Eq.~\ref{eq_hfssum2}), where constructive or destructive interference can occur. To simplify the analysis, we define an effective E1 transition rate based on the total observed Rabi frequency:
\begin{equation}
\mathcal{R}_{\mathrm{E1,eff}}^{(F_e\rightarrow F_g)} = \left(\frac{\Omega^{(F_g \rightarrow F_e)}}{K_{\mathrm{E1}}^{(F_e\rightarrow F_g)}}\right)^2 
\end{equation}

Using this, the actual HFE1 transition rate can be isolated as:
\begin{equation}
\mathcal{R}_{\mathrm{HFE1}}^{(F_e\rightarrow F_g)} = \mathcal{R}_{\mathrm{E1,eff}}^{(F_e\rightarrow F_g)} \left(1 \mp 2\frac{\Omega^{(F_g\rightarrow F_e)}_{\mathrm{E3}}}{\Omega^{(F_g\rightarrow F_e)}} + \left(\frac{\Omega^{(F_g\rightarrow F_e)}_{\mathrm{E3}}}{\Omega^{(F_g\rightarrow F_e)}}\right)^2\right)
\end{equation}

In the limit where $\Omega_{\mathrm{E3}}^{(F_g\rightarrow F_e)} \ll \Omega^{(F_g\rightarrow F_e)}$, the E3 contribution is negligible in the observed Rabi frequency. Therefore this simplifies to $\mathcal{R}_{\mathrm{HFE1}}^{(F_e\rightarrow F_g)} \approx \mathcal{R}_{\mathrm{E1,eff}}^{(F_e\rightarrow F_g)}$, justifying our comparison between experimentally extracted $\mathcal{R}_{\mathrm{E1,eff}}^{(F_e\rightarrow F_g)}$ and theoretical HFE1 rates from Ref.~\cite{dzubaHyperfineInducedElectricDipole2016}. 
The uncertainty arising from this approximation can be estimated as $2\times\Omega^{(F_g \rightarrow F_e)}_{\mathrm{E3}}/\Omega^{(F_g \rightarrow F_e)}$, corresponding to 50\% for $\mathcal{R}_{\mathrm{HFE1}}^{(2\rightarrow3)}$ and 25\% for $\mathcal{R}_{\mathrm{HFE1}}^{(4\rightarrow3)}$.

In summary, the effective E1 transition rate can be calculated using the following equation,
\begin{equation}
   \mathcal{R}_{\mathrm{E1,eff}}^{(F_e\rightarrow F_g)} = \frac{1}{\tau_\mathrm{E3}}\left(\frac{K_{\mathrm{E3}}^{(3\rightarrow 6)}}{K_{\mathrm{E1}}^{(F_e\rightarrow F_g)}}\right)^2 
\left(\frac{\Omega^{(F_g \rightarrow F_e)}}{\Omega^{(3\rightarrow 6)}}\right)^2.
\end{equation}
If the \ybe lifetime $\tau_\mathrm{E3}$ is revised \cite{szSurzhykovPrivComm}, the effective E1 transition rate will update according to this equation.

\section{Discussion on lifetime}

From the individual transition rates for each hyperfine level and for each contribution (E3 and HFE1), we compute the decay rate by summing over all allowed transitions rates, such as:

\begin{equation}\label{eq_decay_rate}
\mathcal{R}^{(Fe)} = \sum_{F_g} \mathcal{R}_{\mathrm{E3}}^{(F_e \rightarrow F_g)} + \sum_{F_g} \mathcal{R}_{\mathrm{HFE1}}^{(F_e \rightarrow F_g)}
\end{equation}

The lifetime of a given hyperfine state is then the inverse of this decay rate:

\begin{equation}\label{eq_lifetime}
\tau^{(F_e)} = \frac{1}{\mathcal{R}^{(F_e)}} = \frac{1}{\sum_{F_g} \mathcal{R}_{\mathrm{E3}}^{(F_e \rightarrow F_g)} + \sum_{F_g} \mathcal{R}_{\mathrm{HFE1}}^{(F_e \rightarrow F_g)}}
\end{equation}

For example, in our case, for each state $\ket{^2F_{7/2}, F_e=2,4,6}$, the lifetime is given by:

\begin{equation}
\begin{aligned}
\tau^{(2)} &= \frac{1}{\mathcal{R}_{\mathrm{E3}}^{(2 \rightarrow 2)} + \mathcal{R}_{\mathrm{E3}}^{(2 \rightarrow 3)} + \mathcal{R}_{\mathrm{HFE1}}^{(2 \rightarrow 2)} + \mathcal{R}_{\mathrm{HFE1}}^{(2 \rightarrow 3)}} \\
\tau^{(4)} &= \frac{1}{\mathcal{R}_{\mathrm{E3}}^{(4 \rightarrow 2)} + \mathcal{R}_{\mathrm{E3}}^{(4 \rightarrow 3)} + \mathcal{R}_{\mathrm{HFE1}}^{(4 \rightarrow 3)}} \\
\tau^{(6)} &= \frac{1}{\mathcal{R}_{\mathrm{E3}}^{(6 \rightarrow 3)}}
\label{eq_tau}
\end{aligned}
\end{equation}

Since the $F_e = 6$ state decays exclusively via the E3 channel to $F_g = 3$, this gives ${\mathcal{R}_{\mathrm{E3}}^{(6 \rightarrow 3)}} = \mathcal{R}^{(6)}=1/\tau^{(6)}$ and we have $\tau^{(6)} = \tau_\mathrm{E3}=1.6(1)$ years from \cite{langeLifetime722021, szSurzhykovPrivComm}.

We also note that the $\mathcal{R}_{\mathrm{HFE1}}^{(2 \rightarrow 2)}$ transition rate has not yet been measured, as the corresponding $\Delta m = 0$ transition is forbidden by selection rules. To access $\Delta m = \pm 1$ transitions, the polarization of the clock laser would need to be oriented orthogonal to the quantization axis. However, this would introduce significant experimental uncertainty due to polarization sensitivity and beam-pointing instability. As a result, we cannot report a reliable value for $\tau^{(2)}$ with the current data.

To approximate the lifetime of the $F_e = 4$ state, we further assume that $\mathcal{R}_{\mathrm{HFE1}}^{(4 \rightarrow 3)} \approx \mathcal{R}_{\mathrm{E1,eff}}^{(4\rightarrow 3)}$. Under this approximation, we can neglect the E3 contribution and estimate the lifetime as:

\begin{equation}
\tau^{(4)} \approx \frac{1}{\mathcal{R}_{\mathrm{E1,eff}}^{(4\rightarrow 3)}}
\label{eq:tau_approx}
\end{equation}

The lifetime for $F_e = 4$ is then $\tau^{(4)}=$ 49(21) days.

\section{Discussion on laser polarization}
Throughout this work, we measure the Rabi frequency driven by a $\pi$-polarized clock laser. 
To validate our results, we first analyze the polarization purity of the clock laser.

\begin{figure}[h]
	\centering\includegraphics[width=0.35\textwidth]{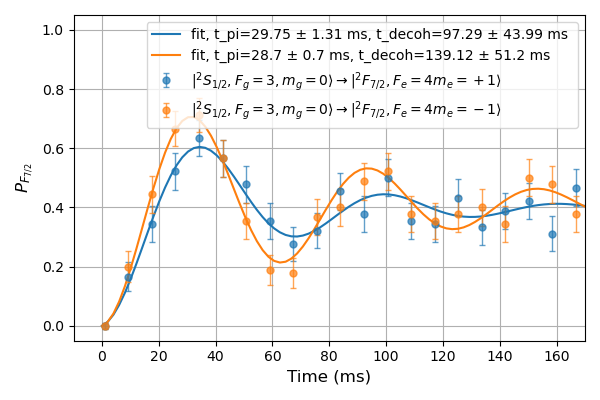}
	\caption{\label{fig:perp_pol}
	Measured Rabi flops of the $\Delta m=\pm1$ transitions, fitted using the JC+EXP2 model. The difference in decoherence time comes from the fact that the Zeeman sensitivities of the $m_e=\pm1$ states are affected by their large quadratic Zeeman effect, leading to stronger decoherence for the $m_e=+1$ Zeeman substate. }
\end{figure}

According to the selection rules for E1 transitions, linearly polarized light can excite the $\Delta m=0$ transition when its polarization is parallel with the magnetic field.
When the polarization is perpendicular to the magnetic field, the $\Delta m=\pm1$ transitions can be excited with equal Rabi frequencies.
If the laser is not purely linearly polarized, a circular polarization component leads to unequal Rabi frequencies for the $\Delta m=\pm1$ transitions.

To verify the polarization purity of the clock laser, we excite the $|^2S_{1/2},F_g=3,m_g=0\rangle \rightarrow |^2F_{7/2},F_e=4,m_e=\pm1\rangle$ transitions.
Note that these transitions are forbidden by the E3 selection rules for our experimental geometry and are therefore driven solely by the HFE1 effect.
The measured Rabi flops are shown in Fig. \ref{fig:perp_pol}.
As the two transitions have the same Rabi frequency within uncertainties, the polarization purity (PER) of the clock laser is better than 14 dB.

For a purely linearly polarized clock laser propagating perpendicular to the quantization axis, a misalignment angle ($\phi$) between the polarization and the quantization axis does not introduce a systematic error in our measurement, since both $\Omega_\mathrm{E1}$ and $\Omega_\mathrm{E3}$ are proportional to $\cos{\phi}$.
Therefore, the laser polarization could only have a marginal effect on our results.

\end{document}